\documentclass[prd,eqsecnum,twocolumn,nofootinbib]{revtex4-2}

\usepackage{amsfonts}
\usepackage{amsmath}
\usepackage{amssymb}
\usepackage{amsthm}
\usepackage{bm}
\usepackage{dcolumn}
\usepackage{epsfig}
\usepackage{graphicx}
\usepackage{graphics}
\usepackage[latin1]{inputenc}
\usepackage{latexsym}
\usepackage{rotating}
\usepackage{xcolor}
\usepackage[mathscr]{euscript}
\usepackage{calligra} \DeclareMathAlphabet{\mathcalligra}{T1}{calligra}{m}{n} \DeclareFontShape{T1}{calligra}{m}{n}{<->s*[2.2]callig15}{}

\begin{document}
\title{Precisely computing bound orbits of spinning bodies around black holes\\ II: Generic orbits}
\author{Lisa V. Drummond}
\affiliation{Department of Physics and MIT Kavli Institute, MIT, Cambridge, MA 02139 USA}
\author{Scott A. Hughes}
\affiliation{Department of Physics and MIT Kavli Institute, MIT, Cambridge, MA 02139 USA}
\begin{abstract}
In this paper, we continue our study of the motion of spinning test bodies orbiting Kerr black holes.  Non-spinning test bodies follow geodesics of the spacetime in which they move.  A test body's spin couples to the curvature of that spacetime, introducing a ``spin-curvature force'' which pushes the body's worldline away from a geodesic trajectory.  The spin-curvature force is an important example of a post-geodesic effect which must be modeled carefully in order to accurately characterize the motion of bodies orbiting black holes.  One motivation for this work is to understand how to include such effects in models of gravitational waves produced from the inspiral of stellar mass bodies into massive black holes.  In this paper's predecessor, we describe a technique for computing bound orbits of spinning bodies around black holes with a frequency-domain description which can be solved very precisely.  In that paper, we present an overview of our methods, as well as present results for orbits which are eccentric and nearly equatorial (i.e., the orbit's motion is no more than $\mathcal{O}(S)$ out of the equatorial plane).  In this paper, we apply this formulation to the fully generic case --- orbits which are inclined and eccentric, with the small body's spin arbitrarily oriented.  We compute the trajectories which such orbits follow, and compute how the small body's spin affects important quantities such as the observable orbital frequencies $\Omega_r$, $\Omega_\theta$ and $\Omega_\phi$.
\end{abstract}
\maketitle

\section{Introduction}
\label{sec:intro}

\subsection{Spinning-body motion around black holes}
\label{sec:emri}

The orbital motion of a spinning test body in a black hole spacetime represents a clean limit of the relativistic two-body problem.  It also is of astrophysical significance as a model for extreme mass-ratio inspirals (EMRIs).  Astrophysical EMRIs consist of stellar-mass compact objects (of mass $\mu$) orbiting a massive black hole (mass $M$).  Such systems are expected to inspiral over their lifetime due to the backreaction of the gravitational waves (GWs) they emit.  If the large black hole is in the mass range $10^5\,M_\odot \lesssim M \lesssim 10^7\,M_\odot$, EMRI waves are expected to radiate in the sensitive frequency band of the planned low-frequency space-based Laser Interferometer Space Antenna (LISA) \cite{eLISA2013,Barausse2020}.  Measurements of EMRI GWs are expected to make possible precision measurements of the properties of the larger black hole \cite{Babak2017} and of the EMRI's astrophysical environment \cite{Kocsis2011, Barausse2014, Derdzinski2019, Bonga2019}.

Enabling such precise measurements will require observers to use accurate waveform models which can match phase with astrophysical signals over the inspiral, both to integrate EMRI signals out of detector noise as well as to facilitate characterizing their sources.  Thanks to their small mass ratio, $\varepsilon \equiv \mu/M\sim10^{-7}\text{--}10^{-4}$, it is natural to use perturbation theory to model EMRIs.  A natural place to begin such models is using Kerr geodesics \cite{Kerr1963} to describe the motion of the smaller body at zeroth order in $\varepsilon$.  We then introduce corrections which encode the nature of ``post-geodesic'' physics that affects the smaller body's motion.  A body travelling on a Kerr geodesic obeys the equation of motion
\begin{equation}
\frac{Dp^{\alpha}}{d\tau} = 0\;,\label{eq:geodesic}
\end{equation}
where $p^{\alpha}$ is the four-momentum of the body, $D/d\tau$ is the covariant derivative computed along the orbit, and $\tau$ is the body's own proper time.  Post-geodesic effects lead to an additional force on the right-hand side of Eq.\ (\ref{eq:geodesic}). In this paper and in our accompanying companion analysis \cite{Paper1}, we study the force that arises when the spin of the small body couples to background spacetime curvature.  The equation describing the small body's motion becomes
\begin{equation}
\frac{Dp^{\alpha}}{d\tau} = f_{S}^{\alpha}\;,
\label{eq:scf}
\end{equation}
which is one of the Mathisson-Papapetrou equations.  We discuss the precise form of the spin-curvature force $f_{S}^{\alpha}$ and the Mathisson-Papapetrou equations in Sec.\ \ref{sec:scc}; see also Sec.\ III\,A of Ref.\ \cite{Paper1} for further details.

This work is a continuation of Ref.\ \cite{Paper1}, which lays out the general framework that we use but presents results only for equatorial or nearly equatorial orbits (``nearly equatorial'' meaning they would be equatorial if the small body were not spinning, but can oscillate by $\mathcal{O}(S)$ out of the equatorial plane due to spin precession effects).  In this paper, we present results for orbits of spinning bodies around black holes with completely generic orbital configurations and spin orientations.  For a discussion of related past work, see Sec.\ I\,B in Ref.\ \cite{Paper1}.

\subsection{Synopsis of our frequency-domain description}
\label{sec:synopsis}

We use a frequency-domain framework to compute the orbits of spinning bodies.  Bound Kerr geodesics naturally lend themselves to this type of treatment as they are characterized by the three coordinate-time frequencies $\hat{\Omega}_r$, $\hat{\Omega}_\theta$ and $\hat{\Omega}_\phi$ related to radial, polar, and axial motions respectively.  This triperiodicity allows for a frequency-domain description of functions which are computed along Kerr orbits:
\begin{equation}
f[\hat r(t),\hat \theta(t)]=\sum_{kn}f_{kn}e^{-in\hat{\Omega}_{r}t} e^{-ik\hat{\Omega}_{\theta}t}\;,
\label{eq:FDgeodesic}
\end{equation}
where $f_{kn}$ are Fourier expansion coefficients.  Notice that the function $f$ we use to illustrate this expansion depends on the orbit's radial coordinate $r$ and polar coordinate $\theta$.  This is common for many relevant functions in our analysis; because the Kerr spacetime is axisymmetric, the coordinate $\phi$ often does not enter the analysis.  Notice also we write certain quantities in (\ref{eq:FDgeodesic}) using a ``hat'' accent, e.g.\ $\hat r(t)$ or $\hat\Omega_r$.  Throughout this paper, we use this accent to denote a quantity which corresponds to a geodesic orbit.  A slight modification to the formulation (\ref{eq:FDgeodesic}) allows us to characterize the properties of spinning-body orbits, as was observed in Ref.\ \cite{Ruangsri2016}.  We describe our frequency-domain formulation in detail in Sec.\ \ref{sec:freqdomdescrip}.

The spin of the small body injects additional harmonic structure into the orbit --- spin precession introduces a new frequency \cite{vandeMeent2019}, which we label $\Omega_{s}$. In addition, the spin of the small body changes the orbital frequencies.  Let us denote the changes relative to an appropriately defined geodesic by $\Omega^S_r$ and $\Omega^S_{\theta}$.  Quantities expanded along a spinning body's orbit, such as the spin-curvature force $f_S^\alpha$, can be written as a Fourier expansion in terms of frequencies $\Omega_{r} = \hat{\Omega}_r + \Omega^S_r$, $\Omega_{\theta} = \hat{\Omega}_{\theta} + \Omega^S_{\theta}$, and $\Omega_{s}$:
\begin{equation}
f[r(t),\theta(t),S^{\mu}(t)]=\sum_{jkn}f_{jkn}e^{-ij\Omega_{s}t}e^{-in\Omega_{r}t}e^{-ik\Omega_{\theta}t}\;.
\label{eq:FDspinorbit}
\end{equation}
Here $S^{\mu}$ is a 4-vector which describes the spin of the small body.  Note that the radial and polar indices $n$ and $k$ both range from $-\infty$ to $\infty$; the spin harmonic index $j$ only varies over the range $j \in [-1,0,1]$.   As with the geodesic expansion (\ref{eq:FDgeodesic}), the frequency-domain expansion (\ref{eq:FDspinorbit}) provides useful machinery for characterizing properties associated with spinning-body orbits.  

We find it convenient to associate each spinning-body orbit with a ``reference'' geodesic.  We thus begin by discussing the parameterization we use for geodesic orbits.  Up to initial conditions, a geodesic is characterized by its semi-latus rectum $p$, its eccentricity $e$ and an inclination angle $I$.  In terms of these parameters, a geodesic's radial and polar motion are parameterized by
\begin{align}
   \hat r & =\frac{p M}{1 + e\cos\hat \chi_r}\;, \ \ \cos\hat \theta = \sin I\cos\hat \chi_\theta\;,
   \label{eq:rtheta_geod}
\end{align}
where the angles $\hat \chi_r$ and $\hat \chi_\theta$ are relativistic versions of ``true anomaly'' angles used in Keplerian orbital dynamics.  Notice that the radial motion oscillates between periapsis at $pM/(1 + e)$ and apoapsis at $pM/(1 - e)$; the polar motion oscillates such that $-\sin I\le \cos\theta \le \sin I$.

Spinning-body orbits have a more ornate structure than geodesics, and in most cases cannot be parameterized in exactly this manner.  An exception is the limit of equatorial orbits in which the small body's spin is aligned with the normal to the orbital plane.  In that case, we set $I = 0^\circ$ or $180^\circ$, and we find we can parameterize the orbit such that it has the same turning points $pM/(1 \pm e)$ as a geodesic orbit.  Note that the motion between turning points differs, however, thanks to the spin-curvature force; see detailed discussion in Secs.\ V and VI of \cite{Paper1}, especially discussion near Eqs.\ (5.16), (5.54), and (6.4).

If the small body's spin is misaligned with the orbit, or the orbit is inclined with respect to the equatorial plane, the libration region varies along the orbit.  These variations couple the radial, polar, and spin precessional motions, complicating the equations of motion, and preventing them from fully separating.  Despite the complications of the libration region's variation, we can constrain the ``purely radial'' motion --- the aspects of the motion which only have harmonics in $\Omega_r$ --- to lie between $pM/(1+e)$ and $pM/(1-e)$.  We can likewise constrain the ``purely polar'' motion, which only has harmonics in $\Omega_{\theta}$, to lie between $-\sin{I}$ and $\sin{I}$.  In this sense, we parameterize the spinning-body orbits with respect to a reference geodesic which has radial and polar turning points precisely at $pM/(1\pm e)$ and $\pm\sin{I}$.  We then compute shifts to important properties of the orbit relative to this reference geodesic. For example, for spinning-body orbits confined entirely to the equatorial plane, we compute shifts to the orbital frequencies relative to geodesic orbits with the same radial turning points; this case is discussed in detail in our companion paper \cite{Paper1}.  In this paper, we further elucidate how reference geodesics are characterized briefly in Sec.\ \ref{sec:spinningbodydescrip}, and in much greater detail in Sec.\ \ref{sec:RefGeodFreqDom}.  In Appendix \ref{sec:referencegeodesics}, we discuss different definitions of reference geodesics (i.e., geodesics ``close to" a corresponding spinning-body orbit) that have been used in the literature.

\subsection{Organization of this paper}

We begin by summarizing the key concepts and notation which underlie our description of spinning-body motion in Sec.\ \ref{sec:mpd}.  We provide a concise review of bound Kerr geodesics in Sec.\ \ref{sec:kerrgeodesics} and we discuss how spin-curvature coupling modifies the equations of motion (relative to a geodesic reference) for a spinning test body in Sec.\ \ref{sec:scc}. In Sec.\ \ref{sec:spinningbodydescrip}, we describe spinning-body orbits qualitatively. This description is then made quantitative as we outline the small-spin perturbative approach (Secs.\ \ref{sec:SpinDev}) and computational framework (Sec.\ \ref{sec:compframework}) that we use to calculate the orbits. 

In Sec.\ \ref{sec:spinbodyfreqdom}, we use a frequency-domain treatment to compute generic orbits, which are both inclined and eccentric, and for which the small body's spin is arbitrarily oriented. We outline the general principles of the frequency-domain description in Sec.\ \ref{sec:freqdomdescrip}. In Secs.\ \ref{sec:circinclalign} and \ref{sec:circinclmisalign}, we focus specifically on spinning-body orbits that are ``nearly circular,'' with aligned spins discussed in Sec.\ \ref{sec:circinclalign} and misaligned spins discussed in Sec.\ \ref{sec:circinclmisalign}.  Nearly circular orbits have an associated reference geodesic that is circular; the orbits have a Boyer-Lindquist coordinate radius that is constant modulo a small variation of $\mathcal{O}(S)$.  In Sec.\ \ref{sec:genericorbits}, we consider the fully generic case, with both arbitrary eccentricities and inclinations. We conclude in Sec.\ \ref{sec:summary} by summarizing our results and outlining plans for related future research. In Appendix \ref{sec:comparWitzany}, we compare our results with an alternative method for computing these frequency shifts presented in Ref.\ \cite{Witzany2019_2}.

As in our companion paper, quite a few of the functions which enter into this analysis are extremely lengthy.  Both because this makes them difficult to read and because the likelihood of introducing errors when typesetting them is high, we provide the explicit formulas for these expressions using a {\it Mathematica} notebook included with this paper's Supplementary Material \cite{SupplementalMaterial}, rather than writing the expressions out in the paper.

Throughout this paper, we work in geometrized units with $G = 1$, $c = 1$.

\section{The motion of a spinning body}
\label{sec:mpd}

In this analysis, we formulate the motion of a spinning body in terms of a nearby ``reference'' geodesic orbit.  In order to introduce important notation and to keep this manuscript self contained, we begin in Sec.\ \ref{sec:kerrgeodesics} with a brief synopsis of Kerr geodesic spacetime.  For a more detailed discussion of these geodesics, see Sec.\ II of the companion paper to this work, Ref.\ \cite{Paper1}, as well as numerous other articles \cite{Schmidt2002, Kraniotis2004, DrascoHughes2004, Hackmann2008, Levin2008, Levin2009, FujitaHikida2009, Hackmann2010, Warburton2013, Rana2019}.  We then summarize the key concepts and equations governing spinning-body orbits in Sec.\ \ref{sec:scc}, with a particular focus on how one describes the parallel transport of an orbiting body's spin angular momentum in Sec.\ \ref{sec:paralleltransport}.

\subsection{Kerr geodesics}
\label{sec:kerrgeodesics}

The metric for the Kerr spacetime in Boyer-Lindquist coordinates is \cite{Boyer1967} 
\begin{align}
ds^2 & =-\left(1-\frac{2r}{\Sigma}\right)\,dt^2+\frac{\Sigma}{\Delta}\,dr^2-\frac{4Mar\sin^2\theta}{\Sigma}dt\,d\phi\nonumber\\
 &+\Sigma\,d\theta^2 +\frac{\left(r^2+a^2\right)^2-a^2\Delta\sin^2\theta}{\Sigma}\sin^2\theta\,d\phi^2,\label{eq:kerrmetric}
\end{align}
where $M$ is the mass and $a$ is the spin parameter $a$ of the black hole and 
\begin{equation}
\Delta =r^2-2Mr+a^2\;,\qquad \Sigma =r^2+a^2\cos^2\theta\;.
\end{equation}
The Kerr geometry possesses  two  Killing vectors $\xi_{t}$ and $\xi_{\phi}$; these Killing vectors yield two constants of motion given by
\begin{align}
\hat E & =-\xi_{t}^{\alpha}u_{\mu}=-\hat u_{t}\;,\\
\hat L_z & =\xi_{\phi}^{\alpha}u_{\mu}=\hat u_{\phi}\;,
\end{align}
where we have normalized these quantities by the mass $\mu$ of the small body.  As mentioned in the Introduction, throughout this paper a quantity with a hat accent, such as $\hat E$, means that it corresponds to a geodesic orbit.

The Kerr spacetime admits an anti-symmetric Killing-Yano tensor which is given by \cite{Penrose1973,Tanaka1996}
\begin{equation}
\mathcal{F}_{\mu\nu}=a\cos\theta\left(\bar e_{\mu}^{1}\bar e_{\nu}^{0}-\bar e_{\mu}^{0}\bar e_{\nu}^{1}\right)+r\left(\bar e_{\mu}^2\bar e_{\nu}^{3}-\bar e_{\mu}^{3}\bar e_{\nu}^2\right)\;,
\end{equation}
where
\begin{align}
\bar e_{\mu}^{0} & =\left[\sqrt{\frac{\Delta}{\Sigma}},0,0,-a\sin^2\theta\sqrt{\frac{\Delta}{\Sigma}}\right]\;,\\
\bar e_{\mu}^{1} & =\left[0,\sqrt{\frac{\Sigma}{\Delta}},0,0\right]\;,\\
\bar e_{\mu}^2 & =\left[0,0,\sqrt{\Sigma},0\right]\;,\\
\bar e_{\mu}^{3} & =\left[-\frac{a\sin\theta}{\sqrt{\Sigma}},0,0,\frac{\left(r^2+a^2\right)\sin\theta}{\sqrt{\Sigma}}\right]\;.
\end{align}
The Kerr metric also admits a Killing tensor $K_{\mu\nu}$ which can be thought of as the ``square'' of $\mathcal{F_{\mu\nu}}$:
\begin{equation}
K_{\mu\nu}=\mathcal{F}_{\mu\alpha}{\mathcal{F}_\nu}^{\alpha}\;.
\end{equation}
The existence of the Killing tensor allows us to define a fourth conserved constant for Kerr geodesic motion \cite{Carter1968} (in addition to the orbiting body's rest mass $\mu$, its energy $\hat E$, and its axial angular momentum $\hat L_z$):
\begin{equation}
\hat K=K_{\alpha\beta}\hat u^{\alpha}\hat u^{\beta}\;.
\end{equation}
This quantity is called the Carter constant.  A related conserved quantity $\hat Q$, also called the Carter constant, is defined by
\begin{align}
\hat Q &=\hat  K - \left(\hat L_z-a\hat E\right)^2\;.
\label{eq:Qdef}
\end{align}
The geodesic equations for the Kerr metric separate in Boyer-Lindquist coordinates as first shown by Carter \cite{Carter1968}, yielding
\begin{align}
\Sigma^2\left(\frac{d\hat r}{d\tau}\right)^2 & = [\hat E(\hat r^2+a^2)-a\hat L_z]^2\nonumber\\
 & \qquad-\hat\Delta[\hat r^2+(\hat L_z-a\hat E)^2+\hat Q]\nonumber\\
 & \equiv R(\hat r)\;,\label{eq:geodr}\\
\Sigma^2\left(\frac{d\hat \theta}{d\tau}\right)^2 & =\hat Q-\cot^2\hat \theta \hat L_z^2-a^2\cos^2\hat \theta(1-\hat E^2)\nonumber\\
 & \equiv\Theta(\hat \theta),\label{eq:geodtheta}\\
\Sigma\frac{d\hat \phi}{d\tau} & =\csc^2\hat \theta \hat L_z+a\hat E\left(\frac{\hat r^2+a^2}{\hat\Delta}-1\right)-\frac{a^2\hat L_z}{\hat\Delta}\nonumber\\
 & \equiv\Phi(\hat r,\hat \theta)\;,\label{eq:geodphi}\\
\Sigma\frac{d\hat t}{d\tau} & =\hat E\left(\frac{(\hat r^2+a^2)^2}{\hat\Delta}-a^2\sin^2\hat \theta\right)\nonumber\\
&\qquad +a\hat L_z\left(1-\frac{\hat r^2+a^2}{\hat\Delta}\right)\nonumber\\
 & \equiv T(\hat r,\hat \theta)\;.\label{eq:geodt}
\end{align}
Introducing a time parameter $\lambda$, called ``Mino time'', such that $d\lambda=d\tau/\Sigma$, allows the radial and polar equations of motion to decouple entirely \cite{Mino2003}. Equations (\ref{eq:geodr}) -- (\ref{eq:geodt}) become
\begin{align}
\left(\frac{d\hat r}{d\lambda}\right)^2 &= R(\hat r)\;,\qquad \left(\frac{d\hat \theta}{d\lambda}\right)^2=\Theta(\hat \theta)\;,
\nonumber\\
\frac{d\hat \phi}{d\lambda} &= \Phi(\hat r,\hat \theta)\;,\qquad \frac{d\hat t}{d\lambda}=T(\hat r,\hat \theta)\;.
\label{eq:geods_mino}
\end{align}
Note that the Kerr geodesic equations can be solved in closed form when they are parameterized using Mino time. The explicit form of these solutions in terms of elliptic functions can be found in Refs. \cite{vandeMeent2019} and \cite{FujitaHikida2009}.

The bounds of Kerr geodesics are defined by a torus with radius ranging between $r_1 \le\hat  r \le r_2$ and polar angle between $\theta_- \le \hat \theta \le (\pi - \theta_-)$. We find it convenient to define turning points $r_1$ and $r_2$ in terms of semi-latus rectum $p$ and eccentricity $e$, according to
\begin{equation}
    r_1 = \frac{pM}{1 - e}\;,\qquad r_2 = \frac{pM}{1 + e}\;,
\end{equation}
and $\theta_-$ can be expressed in terms of inclination angle $I$ where
\begin{equation}
    I = \pi/2 - \mbox{sgn}(L_z)\theta_-\;.
\end{equation}
We can then write $\hat r$ and $\hat \theta$ in terms of these bounds, yielding
\begin{align}
   \hat  r & =\frac{p M}{1 + e\cos\hat \chi_r}\;,
    \label{eq:rdef}\\
    \cos\hat \theta &= \sin I\cos\hat \chi_\theta\;.
    \label{eq:thdef}
\end{align}
The angles $\hat \chi_r$ and $\hat \chi_\theta$ are relativistic generalizations of the ``true anomaly'' angles found in the Keplerian versions of these expressions.

As mentioned in Sec.\ \ref{sec:synopsis}, bound Kerr geodesics are triperiodic \cite{Schmidt2002,DrascoHughes2004,FujitaHikida2009}.  It is convenient to define these frequencies, associated with radial, polar and axial motions, with respect to Mino-time.  We define $\hat \Lambda_{r}$, $\hat \Lambda_{\theta}$, and $\hat \Lambda_{\phi}$ as the radial, polar, and axial Mino-time periods; related to each of these periods is a Mino-time frequency $\hat\Upsilon_{r,\theta,\phi} = 2\pi/\hat\Lambda_{r,\theta,\phi}$.  Because much of our calculation depends on frequency-domain descriptions of geodesic motion, Mino-time frequencies are particularly important.  As shown in Ref.\ \cite{DrascoHughes2004}, Fourier expansions of functions $f(\lambda)=f\left[\hat r(\lambda),\hat \theta(\lambda)\right]$ evaluated along Kerr orbits can be written
\begin{equation}
f = \sum_{k=-\infty}^\infty\sum_{n = -\infty}^\infty f_{kn}e^{-i\left(k\hat \Upsilon_{\theta}+n\hat \Upsilon_{r}\right)\lambda}
\end{equation}
where the Fourier coefficient $f_{kn}$ is straightforwardly computed using
\begin{widetext}
\begin{equation}
f_{kn}=\frac{1}{\hat \Lambda_{r}\hat \Lambda_{\theta}}\int_{0}^{\hat \Lambda_{r}}\int_{0}^{\hat \Lambda_{\theta}}f\left[\hat r(\lambda_{r}),\hat \theta(\lambda_{\theta})\right]e^{ik\hat \Upsilon_{\theta}\lambda_\theta}e^{in\hat \Upsilon_{r}\lambda_r}d\lambda_{\theta}d\lambda_{r}\;.
\end{equation}
\end{widetext}
The quantities $\hat\Upsilon_\phi$ and $\hat\Gamma$ are defined as the orbit averages of the functions $\Phi( {\hat r}, {\hat \theta})$ and $T( {\hat r},  {\hat \theta})$ in Eq.\ (\ref{eq:geods_mino}), where $f_{00}$ is the orbit-average of the function $f[{\hat r(\lambda)}, {\hat \theta(\lambda)}]$. The quantity $\hat\Gamma$ is used to convert between the  Boyer-Lindquist coordinate-time frequencies $\hat\Omega_{r,\theta,\phi}$ and Mino-time frequencies $\hat\Upsilon_{r,\theta,\phi}$, via
\begin{equation}
\hat\Omega_{r,\theta,\phi}=\frac{\hat\Upsilon_{r,\theta,\phi}}{\hat\Gamma}\;.
\end{equation}
See Sec.\ II\,C of Ref.\ \cite{Paper1} for further detail on the frequency-domain description of geodesic motion. In this article, Sec.\ \ref{sec:coordinatetime} provides a prescription for computing the coordinate-time analogues of Mino-time frequencies associated with spinning-body orbits.

\subsection{Spin-curvature coupling}
\label{sec:scc}
The motion of a spinning body is governed by the Mathisson-Papapetrou equations \cite{Papapetrou1951, Mathisson2010,Mathisson2010G_2,Dixon1970}
\begin{align}
\frac{Dp^{\alpha}}{d\tau} & =-\frac{1}{2}{R^\alpha}_{\,\nu\lambda\sigma}u^{\nu}S^{\lambda\sigma}\;,\label{eq:mp1}\\
\frac{DS^{\alpha\beta}}{d\tau} & =p^{\alpha}u^{\beta}-p^{\beta}u^{\alpha}\;,\label{eq:mp2}
\end{align}
where $S^{\alpha \beta}$ is the spin tensor and the 4-momentum $p^\mu$ is given by
\begin{equation}
p^{\alpha}=\mu u^{\alpha}-u_{\gamma}\frac{DS^{\alpha\gamma}}{d\tau}\;.\label{eq:momvel}
\end{equation}
To close the system of equations (\ref{eq:mp1}) -- (\ref{eq:mp2}), we need an additional constraint known as the spin supplementary condition (SSC). A commonly used SSC is the Tulczyjew SSC,
\begin{equation}
p_{\alpha}S^{\alpha\beta}=0\;,\label{eq:TD}
\end{equation}
which we employ throughout this analysis \cite{Tulczyjew1959}.  We define the spin vector in terms of the spin tensor through  \cite{Kyrian2007}
\begin{equation}
S^{\mu}=-\frac{1}{2\mu}{\epsilon^{\mu\nu}}_{\alpha\beta}p_{\nu}S^{\alpha\beta}\;,
\label{eq:spinvec}
\end{equation}
where
\begin{equation}
\epsilon_{\alpha\beta\gamma\delta}=\sqrt{-g}[\alpha\beta\gamma\delta]\;,
\end{equation}
$\sqrt{-g}$ is the metric determinant, and $[\alpha\beta\gamma\delta]$ is the totally antisymmetric symbol. 

One can define two conserved quantities associated with these equations: the energy and axial angular momentum per unit mass. These are given by
\begin{align}
E^S & = -u_t+\frac{1}{2\mu}\partial_{\beta}g_{t\alpha}S^{\alpha\beta} \label{eq:Espin},\\ 
L_z^S & = u_{\phi}-\frac{1}{2\mu}\partial_{\beta}g_{\phi\alpha}S^{\alpha\beta}.\label{eq:Lspin}
\end{align}
respectively. The magnitude of the spin vector $S$ is another constant of motion, given by
\begin{equation}
S^2=S^{\alpha}S_{\alpha}=\frac{1}{2}S_{\alpha\beta}S^{\alpha\beta}\;.\label{eq:smag}
\end{equation}
The magnitude $S$ can then be defined in terms of a dimensionless spin parameter $s$,
\begin{equation}
S=s\mu^2\;.\label{eq:dimesionless}
\end{equation}
If the smaller body is itself a Kerr black hole, then $0 \le s \le 1$.  In addition, $p^\mu p_\mu=-\mu^2$ is constant along the worldline of the orbiting body to linear order in $S$. Finally, an analogue of the Carter constant is conserved at linear order in $S$ and is given by \cite{Rudiger1981}
\begin{equation}
K^S=K_{\alpha\beta}u^\alpha u^\beta+\delta\mathcal{C}^S\;,
\label{eq:Kspin}
\end{equation}
where 
\begin{equation}
\delta\mathcal{C}^S= -\frac{2}{\mu}\hat u^{\mu}S^{\rho\sigma}\left( {\mathcal{F}^\nu}_{\sigma}\nabla_{\nu}\mathcal{F}_{\mu \rho } - {\mathcal{F}_\mu}^\nu\nabla_{\nu}\mathcal{F}_{\rho\sigma}\right)\;.
\label{eq:Cspin}
\end{equation}

Using the Tulczyjew SSC in Eq.\ (\ref{eq:TD}), we can deduce that $p^\alpha = \mu u^\alpha + \mathcal{O}(S^2)$.  Examining the motion to leading order in the small body's spin, we therefore have
\begin{equation}
p^{\alpha}=\mu u^{\alpha}\;,
\label{eq:momvelfirstorder}
\end{equation}
i.e., 4-velocity and 4-momentum are parallel at this order.
Accordingly, Eqs.\ (\ref{eq:mp1}) -- (\ref{eq:mp2}) now become
\begin{align}
\frac{Du^{\alpha}}{d\tau} & =-\frac{1}{2\mu}{R^\alpha}_{\,\nu\lambda\sigma}u^{\nu}S^{\lambda\sigma}\;,
\label{eq:mp1linear}\\
\frac{DS^{\alpha\beta}}{d\tau} &= 0\;,\label{eq:mp2linear}
\end{align}
to leading order in small-body spin. Once we have linearized in spin, we can write the small body's 4-velocity as
\begin{equation}
u^{\alpha}=\hat{u}^{\alpha}+u_{S}^{\alpha}\;,
\label{eq:4vellinear}
\end{equation}
where $\hat u^\alpha$ solves the geodesic equation. As first noted in Sec.\ \ref{sec:synopsis}, the hat accent denotes quantities that are evaluated along a geodesic with 4-velocity $\hat u^\alpha$; $u_S^\alpha$ then denotes the $\mathcal{O}(S)$ correction to the 4-velocity.  Equation (\ref{eq:spinvec}) becomes
\begin{equation}
S^{\mu}=-\frac{1}{2}{\epsilon^{\mu\nu}}_{\alpha\beta}\hat{u}_{\nu}S^{\alpha\beta}\;,\label{eq:spinveclinear}
\end{equation}
once we have linearized in $S$.  Equivalently, we can write
\begin{equation}
S^{\alpha\beta}=\epsilon^{\alpha\beta\mu\nu}\hat{u}_{\mu}S_{\nu}\;.\label{eq:spinveclinear2}
\end{equation}

\subsection{Parallel transport in Kerr}
\label{sec:paralleltransport}
When we combine Eqs.\ (\ref{eq:mp2linear}) and (\ref{eq:spinveclinear2}), we obtain
\begin{equation}
\frac{DS^{\mu}}{d\tau}=0\;,
\label{eq:mp2linear2}
\end{equation}
which means that the spin vector is parallel transported at this order. Parallel transport of a vector in the Kerr spacetime has a closed form solution presented in Ref.\ \cite{vandeMeent2019} which builds on the tetrad formulation introduced by Marck \cite{Marck1983, Marck1983_2, Kamran1986}. Following Ref.\ \cite{vandeMeent2019}, we outline the procedure for constructing tetrad legs $\{e_{0\alpha},\tilde{e}_{1\alpha},\tilde{e}_{2\alpha},e_{3\alpha}\}$. First, we observe that, by definition, $\hat{u}^\mu$ is parallel-transported along a geodesic worldline. We let $ e_{0\alpha} = \hat u_\alpha$ be the first leg of the tetrad. Next, we define the vector
\begin{equation}
   \hat{\mathcal{L}}^\nu=\mathcal{F}^{\mu\nu}\hat u_{\mu}\;,
    \label{eq:orbangmomdef}
\end{equation}
which we call the orbital angular momentum 4-vector. This vector is also parallel transported in Kerr, so we define $e_{3\alpha} =\hat{\mathcal{L}}_\alpha(\lambda)/\sqrt{\hat K}$ as the fourth leg of the tetrad.  It's worth noting that other sources (e.g., \cite{vandeMeent2019, Witzany2019_2}) define $\mathcal{L}^\nu$ with the contraction on the other index of $\mathcal{F}^{\mu\nu}$.  Because of the antisymmetry of the Killing-Yano tensor, this introduces an overall minus sign.  As discussed in our companion paper \cite{Paper1}, this sign flip insures that the angular momentum has the right components in the equatorial limit (in particular, that $\hat{\mathcal{L}}^\theta \propto -\hat L_z$).  We have found that this convention is needed for our results to agree with past post-Newtonian results.

We next define $\tilde{e}_{1\alpha}$ and $\tilde{e}_{2\alpha}$ by constructing two vectors which lie in the plane orthogonal to $e_{0\alpha}$ and $e_{3\alpha}$; explicit expressions for $\tilde{e}_{1\alpha}$ and $\tilde{e}_{2\alpha}$ are given in Eqs.\ (50) and (51) of Ref.\ \cite{vandeMeent2019}. We let
\begin{align}
    e_{1\alpha} &= \cos\psi_p(\lambda)\,\tilde{e}_{1\alpha} + \sin\psi_p(\lambda)\,\tilde{e}_{2\alpha}
    \label{eq:tetradleg1}\;,\\
    e_{2\alpha} &= -\sin\psi_p(\lambda)\,\tilde{e}_{1\alpha} + \cos\psi_p(\lambda)\,\tilde{e}_{2\alpha}\;,
    \label{eq:tetradleg2}
\end{align}
where we define $\psi_p(\lambda)$ such that
\begin{equation}
\frac{d\psi_p}{d\lambda}=\sqrt{\hat{K}}\left(\frac{(r^2+a^2)\hat{E}-a\hat{L}_z}{\hat{K}+r^2}+a\frac{\hat{L}_z-a(1-z^2)\hat{E}}{\hat{K}-a^2z^2}\right).
\label{eq:precphaseeqn}
\end{equation}
By construction, we have now obtained tetrad legs $\{e_0,e_1,e_2,e_3\}$ that are orthogonal, normalized and parallel transported along geodesics \cite{Marck1983, Marck1983_2, vandeMeent2019}. As mentioned above, a closed form solution of Eq.\ (\ref{eq:precphaseeqn}) is presented in Ref.\ \cite{vandeMeent2019} with the form
\begin{equation}
    \psi_p(\lambda) = \Upsilon_s\lambda + \psi_r(\Upsilon_r\lambda) + \psi_\theta(\Upsilon_\theta\lambda)\;,
    \label{eq:precphasesol}
\end{equation}
where $\Upsilon_s$ is the Mino-time frequency of the precession of this tetrad along the geodesic. We let $\Lambda_s=2\pi/\Upsilon_s$ be the Mino-time precession period. The {\tt KerrGeodesics} package of the Black Hole Perturbation Toolkit \cite{Kerrgeodesics} includes code for computing these tetrad legs and $\Upsilon_s$.

In general, the spin vector of the small body can be written
\begin{equation}
    S_\alpha = S^0 e_{0\alpha}(\lambda) + S^1 e_{1\alpha}(\lambda) + S^2 e_{2\alpha}(\lambda) + S^3 e_{3\alpha}(\lambda)\;,
    \label{eq:spinvectetrad}
\end{equation}
where $\{S^0, S^1, S^2, S^3\}$ are all constants.  The Tulczyjew SSC in Eq.\ (\ref{eq:TD}) requires that $S^0 = 0$.  The component $S^3 \equiv s_\parallel$ describes components of the small body's spin aligned or antialigned with the orbital angular momentum; $S^1$ and $S^2$ are components in the orbital plane, perpendicular to the direction of orbital angular momentum.  Using the dimensionless spin parameter $0 \le s \le 1$ defined in Eq.\ (\ref{eq:dimesionless}), we can write $S_\alpha$ as 
 \begin{equation}
    S_\alpha = \mu^2\bigl(s_\perp\cos\phi_s\,e_{1\alpha} + s_\perp\sin\phi_s\,e_{2\alpha} + s_\parallel\,e_{3\alpha}\bigr)\;.
    \label{eq:Smisalign1}
\end{equation}
where $s = \sqrt{s_\perp^2 + s_\parallel^2}$, and $\phi_s$ describes the orientation of the spin components in the orbital plane. In terms of $\tilde{e}_{1\alpha}$ and $\tilde{e}_{2\alpha}$, we have
  \begin{align}
    S_\alpha &= \mu^2\biggl[s_\perp\Bigl(\cos(\phi_s + \psi_p)\tilde{e}_{1\alpha} + \sin(\phi_s + \psi_p)\tilde{e}_{2\alpha}\Bigr)
    \nonumber\\
    &\qquad + s_\parallel e_{3\alpha}\biggr]\;.
    \label{eq:Smisalign2}
\end{align}
 The small body's spin precesses only when either $S^1$ or $S^2$ are non-zero and hence $s_\perp\neq0$. In this case, the frequency-domain description of $S_\alpha$ includes harmonics of the spin-precession frequency $\Upsilon_s$. If $s_\perp=0$, the spin vector does not precess.

\section{Generic spinning-body orbits: general principles}
\label{sec:spinningbodyorbits}

We begin our discussion of spinning-body motion by presenting a qualitative overview of their orbits and the parameterizations used to describe them (Sec.\ \ref{sec:spinningbodydescrip}).  In Sec.\ \ref{sec:SpinDev}, we then discuss spin-induced deviations to geodesic trajectories and orbital quantities. In Sec.\ \ref{sec:compframework}, we present the mathematical framework we use to compute spinning-body orbits.

\subsection{Characteristics of spinning-body orbits}
\label{sec:spinningbodydescrip}

Spinning-body orbits generally tend to be qualitatively distinct from geodesic orbits.  The most obvious difference is the introduction of harmonics at frequency $\Upsilon_s$, which appear when $s_\perp \ne 0$ due to the precession of the spin vector. However, even when $s_\perp = 0$ so that there is no spin precession, the libration range can vary over the course of the orbit due to harmonics of both $\Upsilon_r$ and $\Upsilon_\theta$.  Unlike the geodesic orbits given in Eqs.\ (\ref{eq:geods_mino}), the radial and polar motions of a spinning body do not fully separate thanks to their coupling via the variations in orbit's libration region.  Instead, the radial turning points are functions of $\theta$ and $\psi_p$, while the polar turning points are functions of $r$ and $\psi_p$; see Ref.\ \cite{Witzany2019_2} for explicit analytic expressions for turning point corrections.

Because bound geodesics have turning points that are fixed for the duration of the orbit, we cannot in general find a geodesic with the same turning points as a given spinning-body orbit.  We find, however, that aspects of the motion which are totally described by harmonics of the radial frequency $\Omega_r$ do in fact have fixed radial turning points; we call this the ``purely radial'' portion of the orbit.  Likewise, aspects of the motion which are totally described by harmonics of the polar frequency $\Omega_\theta$ have fixed polar turning points; we call this the ``purely polar'' orbital motion.  With this in mind, we define the reference geodesic as the geodesic that has the same radial turning points as the purely radial part of the spinning-body orbit, and that has the same polar turning points as the purely polar part of that orbit.  Aspects of the motion which cannot be written as purely radial or purely polar describe variations in the orbit's turning points, and are incorporated into functions which combine the radial, polar, and precessional frequencies.  Section \ref{sec:RefGeodFreqDom} expands on this idea, providing computational detail; see also Appendix \ref{sec:referencegeodesics} for discussion of alternative mappings between geodesic and spinning-body orbits used in the literature.

The simplest case is an equatorial orbit with aligned spin, so that $s_\perp = 0$.  In this case, an orbit's radial and polar motion can be parameterized as
\begin{equation}
r =\frac{p M}{1 + e\cos\chi_r}\;,\ \ \theta=\frac{\pi}{2}\;.
\end{equation}
This constrains the radial motion to the interval $p/(1+e)\leq r \leq p/(1-e)$, exactly as for geodesic motion.  Note, however, that the true anomaly angle $\chi_r$ for the spinning-body orbit is not the same as the geodesic true anomaly $\hat\chi_r$: there is a shift in the radial frequency from $\hat\Upsilon_r$ to $\Upsilon_r = \hat\Upsilon_r + \Upsilon^S_r$, as well as a shift to an oscillating contribution to this angle.

For misaligned spin, with $s_\perp \neq 0$, the spin vector precesses and truly equatorial orbits do not exist.  However, we can find ``nearly equatorial'' orbits which oscillate $\mathcal{O}(S)$ out of the equatorial plane.  For the nearly equatorial orbits, we can still parameterize the radial motion in the same way as a geodesic, but there are adjustments to the polar libration range due to the spin precession.  The turning points of the polar motion then depend on the spin precession phase $\psi_p$.  We write nearly equatorial orbits in the form
\begin{equation}
    r = \frac{pM}{1 + e\cos\chi_r}\;, \ \ \theta=\frac{\pi}{2} + \delta \vartheta_S\;\label{eq:rparam2},
\end{equation}
where the angle $\delta\vartheta_S$ describes the $\mathcal{O}(S)$ librations in polar angle.  We investigate these orbits in detail in our companion paper \cite{Paper1}.

Spinning-body orbits which are inclined with respect to the equatorial plane cannot be parameterized in the same way as geodesics even when $s_\perp = 0$.  Inclined orbits with aligned spin (i.e., with $s_\perp = 0$) that are $\mathcal{O}(S)$ away from circular --- ``nearly circular'' orbits --- can be parameterized using
\begin{align}
r &=pM + \delta\mathcalligra{r}_S \;, \label{eq:paramcircinclr0}
\\
\cos\theta & =\sin I\cos \chi_{\theta}\;.\label{eq:paramcircincltheta0}
\end{align}
The polar motion in this parameterization is the same as that for an inclined geodesic orbit (bearing in mind that the true anomaly angle $\chi_\theta$ differs from the true anomaly $\hat\chi_\theta$ that describes a geodesic), but the radial motion includes a function $\delta\mathcalligra{r}_S$ which accounts for oscillations in the radial libration region due to spin-curvature coupling.  This form is discussed in detail in Section \ref{sec:circinclalign}.

For nearly circular orbits with misaligned spin, the radial turning points depends on both $\theta$ and $\psi_p$; the polar turning points depend on $r$ and $\psi_p$.  The orbits in this case are described by
\begin{align}
r & =pM + \delta\mathcalligra{r}_S\;, 
\label{eq:paramcircinclprecessr0}
\\ \cos \theta &=\sin I\cos \chi_\theta + \delta \mathcalligra{z}_S\;.\label{eq:paramcircinclprecesstheta0}
\end{align}
The function $\delta\mathcalligra{z}_S$ accounts for variations in the $\cos\theta$ libration region.  This parameterization can be written as a variation in polar angle:
\begin{equation}
    \delta\mathcalligra{z}_S = - \delta\vartheta_S\sqrt{1 - \sin^2I\cos^2\chi_\theta}\;.
\end{equation}
This relationship is most useful for nearly equatorial orbits which have $\sin I = 0$, for which $\delta\mathcalligra{z}_S = -\delta\vartheta_S$.  Circular, inclined orbits with misaligned spin are discussed in detail in Sec.\ \ref{sec:circinclmisalign}. 

Finally, in the fully generic case when the orbit is eccentric and inclined with arbitrarily oriented spin, the parameterization we use has the form
\begin{align}
    r & =\frac{p M}{1 + e\cos\chi_r} + \delta\mathcalligra{r}_S\;,\label{eq:rgenparam0} \\
    \cos \theta &=\sin I\cos\chi_\theta + \delta\mathcalligra{z}_S\;.\label{eq:thetagenparam0}
\end{align}
This case is discussed in detail in Sec.\ \ref{sec:genericorbits}. 

\subsection{Perturbative framework for the motion of spinning bodies}
\label{sec:SpinDev}

In Eq.\ (\ref{eq:dimesionless}), we defined a dimensionless spin parameter $s$ which satisfies $0 \le s \le 1$ if the small body is itself a Kerr black hole.  The magnitude of the small body's spin is then $S \le \mu^2$, and so linear-in-spin effects are quadratic in the system's mass ratio.  In what follows, we neglect terms in our equations that are $\mathcal{O}(S^2)$ or higher, as such terms are negligible for the extreme mass ratio systems we are interested in.  Our approach thus hinges on the use of perturbation theory in the mass ratio. 

With a linear-in-spin analysis in mind, it is possible to write the small body's trajectory as 
\begin{equation}
    x^\alpha(\lambda) = \hat x^\alpha(\lambda) + \delta x_S^\alpha(\lambda)\;.
    \label{eq:trajectoryshift}
\end{equation}
Here, $\hat x^\alpha(\lambda)$ is the  trajectory of a geodesic, and $\delta x_S^\alpha(\lambda)$ is the $\mathcal{O}(S)$-deviation from the geodesic trajectory due to the spin of the small body. Similarly, as defined in Eq.\ (\ref{eq:4vellinear}), we can write
\begin{equation}
u^{\alpha}=\hat{u}^{\alpha}+u_{S}^{\alpha}\;.
\label{eq:4vellinear2}
\end{equation}
Observe, however, that the periodic motions which contribute to $\hat{x}^\alpha(\lambda)$ in general have different periods than the ones which contribute to $x^\alpha(\lambda)$.  Using Eq.\ (\ref{eq:trajectoryshift}), we therefore expect $\delta x^\alpha_S$ to contain secular terms which grow without bound.  This means that $\delta x_S^\alpha(\lambda)$ as defined in Eq.\ (\ref{eq:trajectoryshift}) cannot easily be characterized using a frequency-domain description

For this reason, we do not directly use the form Eq.\ (\ref{eq:trajectoryshift}) when we evaluate spinning-body orbits in Sec.\ \ref{sec:spinbodyfreqdom}. We instead parameterize spinning-body orbits using amplitude-phase variables, where the frequency shift is incorporated into the parameterization; see Eqs.\ (\ref{eq:rgenparam1}) -- (\ref{eq:thetagenparam1}) and surrounding text.  These variables are either periodic or constant and do not contain secularly growing terms; they can be described using Fourier expansions as outlined in Sec.\ \ref{sec:freqdomdescrip}.  Once we have solved for the frequency shifts and other unknowns, it is then possible to compute radial and polar spin corrections $\delta r_S$ and $\delta \theta_S$, whose explicit forms in terms of the amplitude-phase variables are given by Eqs.\ (\ref{eq:secularr}) and (\ref{eq:seculartheta}).

One of our goals is to compute corrections relative to geodesic motion of important quantities associated with the orbit. Such quantities include the constants of motion, which we write in the form
\begin{align}
\mathcal{X}^S = \hat{\mathcal{X}} + \delta\mathcal{X}^S\;,
 \end{align}
where $\mathcal{X} \in [E, L_z, K, Q]$.  Here $\hat{\mathcal{X}}$ is the quantity associated with the reference geodesic and $\delta\mathcal{X}^S$ is the correction required when we include the spin of the orbiting body. Explicitly, the leading-order-in-spin corrections to the energy $\delta E^S$ and axial angular momentum $\delta L_z^S$ are defined by
\begin{align}
E^{S}=\hat E + \delta E^S\;,\; L_{z}^{S}=\hat{L}_z + \delta L_z^S\;.\label{eq:deltaLspin}
\end{align}
where $E^{S}$ and $L_{z}^{S}$ are given by Eqs.\ (\ref{eq:Espin}) and (\ref{eq:Lspin}). Similarly, the first order in spin correction to $K$ is defined by
\begin{equation}
K^{S}=\hat{K}+\delta K^S\;, \label{eq:deltaKspin}\\
\end{equation}
where
\begin{align}
    \delta K^S &= 2K_{\alpha\beta}\hat u^\alpha u^\beta_S + \delta r_S \partial_r K_{\alpha\beta}\hat u^\alpha\hat u^\beta + \delta \theta_S \partial_\theta K_{\alpha\beta}\hat u^\alpha\hat u^\beta
    \nonumber\\
    & + \delta\mathcal{C}^S\;.
    \label{eq:deltaKspin2}
\end{align}
and where $\delta\mathcal{C}^S$ is given by Eq.\ (\ref{eq:Cspin}). Finally, using Eq.\ (\ref{eq:Qdef}), we can obtain the first-order shift in $Q$:
\begin{equation}
    \delta Q^S = \delta K^S - 2(\hat L_z - a\hat E)(\delta L^S_z - a\delta E^S)\;.
    \label{eq:deltaQspin}
\end{equation}

The spin of the small body also introduces corrections to the fundamental frequencies of the orbit, which we write in the form
\begin{align}
    \Upsilon_x &= \hat\Upsilon_x + \Upsilon^S_x\;, \ \ \Gamma = \hat\Gamma + \Gamma^S\;, \label{eq:UpsilonGammashifts}
\end{align}
where $x \in [r,\theta, \phi]$. As discussed in Sec.\ \ref{sec:paralleltransport}, the spin of the small body also introduces the spin-precession frequency $\Upsilon_{s}$ into the motion, meaning that orbits of spinning bodies can generally be described using Mino-time Fourier expansions with harmonics of frequencies $\hat{\Upsilon}_{r}+ \Upsilon^S_r$, $\hat{\Upsilon}_{\theta}+\Upsilon^S_{\theta}$ and $\Upsilon_{s}$. This frequency-domain approach is what we will use in Sec.\ \ref{sec:spinbodyfreqdom} to compute properties of spinning-body orbits.

\subsection{Computing spinning-body orbits}
\label{sec:compframework}
We now outline the explicit mathematical framework we use to compute the modification to the small body's trajectory arising from the spin-curvature interaction. Eq.\ (\ref{eq:mp1linear}) is the governing equation for the spinning-body orbits discussed in this work. We repeat this equation below:
\begin{equation}
    \frac{Du^\alpha}{d\tau} = -\frac{1}{2\mu}{R^\alpha}_{\nu\lambda\sigma}u^\nu S^{\lambda\sigma} \equiv f^\alpha_S/\mu\;.
\end{equation}
We define the right-hand side of this equation to be the spin-curvature force $f^\alpha_S$. When we expand the covariant derivative, we have
\begin{equation}
    \frac{du^\alpha}{d\tau} + {\Gamma^\alpha}_{\beta\gamma} u^\beta u^\gamma = f^\alpha_S/\mu\;,
    \label{eq:forcedgeodesic1}
\end{equation}
where ${\Gamma^\alpha}_{\beta\gamma}$ is the Christoffel connection for the Kerr spacetime. We find it convenient to perform all our calculations in Mino-time, so we define 
\begin{equation}
    U^\alpha \equiv \frac{dx^\alpha}{d\lambda} = \Sigma u^\alpha\;,
    \label{eq:Udef}
\end{equation}
where the 4-velocity is $u^\alpha = dx^\alpha/d\tau$ and Mino-time is defined by $d/d\lambda = \Sigma d/d\tau$.  Now that we have defined $U^\alpha$ by Eq.\ (\ref{eq:Udef}), we multiply Eq.\ (\ref{eq:forcedgeodesic1}) by $\Sigma^2$, yielding
\begin{equation}
    \frac{dU^\alpha}{d\lambda}+\Pi^\alpha = F_S^\alpha/\mu \;,
    \label{eq:forcedgeodesic2}
\end{equation}
where
\begin{equation}
 F_S^\alpha\equiv\Sigma^2 f_S^\alpha\;, \ \ 
  \Pi^\alpha\equiv-\frac{U^\alpha}{\Sigma}\frac{d\Sigma}{d\lambda} + {\Gamma^\alpha}_{\beta\gamma}U^\beta U^\gamma\;.
\end{equation}
Consider Eq.\ (\ref{eq:forcedgeodesic2}) component by component.  We start with the axial and temporal components of the 4-velocity.  Begin by writing $u_t$ and $u_\phi$ as
\begin{align}
u_t=-\hat{E}+ u_t^S, \ \ \ \
u_{\phi}=\hat{L}_z+ u_{\phi}^S,\label{eq:utuphi}
\end{align}
where $u^S_{t,\phi}=\mathcal{O}(S)$.  Combining the axial and temporal components of Eq.\ (\ref{eq:mp1linear}) yields two equations of the form
\begin{align}
\frac{d u^S_{\phi}}{d\lambda}= \mathcal{R}_\phi\;, \ \ \ \frac{d u^S_{t}}{d\lambda}= \mathcal{R}_t\;, \label{eq:Reqs}
\end{align}
where $\mathcal{R}_\phi$ and $\mathcal{R}_t$ are functions of known geodesic quantities.  For the case of nearly equatorial orbits, these functions are given in Eqs.\ (5.13) and (5.14) of our companion paper \cite{Paper1}; for the general case, they are among the functions which we include in the supplementary {\it Mathematica} notebook which accompanies this paper.  Using Eqs.\ (\ref{eq:Reqs}), we can then solve for $u^S_{t}$ and $u^S_{\phi}$.

Turn next to the radial and polar components of Eq.\ (\ref{eq:forcedgeodesic2}), which we write
\begin{align}
    \frac{d^2r}{d\lambda^2} &+ \Pi^r = F_S^r\;, \label{eq:forcer}\\  
    \frac{d^2\theta}{d\lambda^2}&+ \Pi^\theta = F_S^\theta\;.\label{eq:forcetheta}
\end{align}
We solve Eqs.\ (\ref{eq:forcer}) and (\ref{eq:forcetheta}) by linearizing in spin and expanding in the frequency domain.  In addition, we preserve the norm of the 4-velocity along the orbit, requiring that
\begin{equation}
u^{\alpha}u_{\alpha}=-1\;.
\label{eq:4vellinear3}
\end{equation}
We linearize Eq.\ (\ref{eq:4vellinear3}) in spin, and expand in the frequency domain.  Our full frequency-domain treatment of spinning-body orbits is discussed in detail in Sec.\ \ref{sec:spinbodyfreqdom}.

\section{Generic spinning-body orbits: Frequency-domain treatment}
\label{sec:spinbodyfreqdom}

We now compute spinning-body orbits which have arbitrary eccentricity and inclination, using a frequency-domain treatment of the spinning body's motion.  In our companion paper, Ref.\ \cite{Paper1}, we described equatorial and nearly equatorial spinning-body orbits in detail.  In that work, we used essentially the same frequency-domain techniques to study equatorial (aligned spin) and nearly equatorial (misaligned spin) orbits with arbitrary eccentricity.  We now extend this technique to encompass orbits that have any orbital inclination, not only those that are within polar angles $\mathcal{O}(S)$ of the equatorial plane.

\subsection{Frequency-domain description}
\label{sec:freqdomdescrip}

Writing quantities defined on a spinning body's orbit in expansions of the form
\begin{equation}
f(\lambda)=\sum_{j=-1}^{1}\sum_{n,k=-\infty}^{\infty}f_{jnk}e^{-i\left(j \Upsilon_{s}+n \Upsilon_{r}+k \Upsilon_{\theta}\right)\lambda}\;,\label{freqexpand}
\end{equation} 
allows us to compute orbits to a high level of precision.  The Fourier coefficient $f_{jnk}$ is defined by 
\begin{align}
f_{jnk}&= \frac{1}{\Lambda_r \Lambda_{\theta} \Lambda_{s}}\nonumber \int_0^{\Lambda_r}\int_0^{\Lambda_{\theta}}\int_0^{\Lambda_s}f(\lambda_r,\lambda_\theta,\lambda_s)\\ &\times e^{i\left(j \Upsilon_{s}\lambda_s+n \Upsilon_{r}\lambda_r+k \Upsilon_{\theta}\lambda_\theta\right)} \ d \lambda_r d \lambda_{\theta} d \lambda_{s}\;.
\end{align}
The techniques we describe below allow us to precisely compute a spinning body's orbital frequencies $\Upsilon_r$ and $\Upsilon_{\theta}$ for fully generic orbits.  As discussed and defined in Eq.\ (\ref{eq:UpsilonGammashifts}), we treat these frequencies as ``spin shifted'' relative to the the radial and polar frequencies of a reference geodesic, writing $\Upsilon_r = \hat\Upsilon_r + \Upsilon_r^S$ and $\Upsilon_\theta = \hat\Upsilon_\theta + \Upsilon_\theta^S$.
 
\subsubsection{Generalities}
\label{sec:generalities}

We first examine the $t$ and $\phi$ components of the 4-velocity.  The frequency-domain expansion allows us to solve the axial and temporal components of Eq.\ (\ref{eq:mp1linear}), which we write explicitly in the form shown in Eqs.\ (\ref{eq:Reqs}).  To do this, we expand $u_t^S$ and $u_{\phi}^S$ as\footnote{Note that if the function we are Fourier expanding already has a subscript, we use a comma to denote the specific Fourier mode. For example, $u_{t,1,0,-1}^S$ is the $j=1$, $n=0$, $k=-1$ harmonic of function $u_t^S$.}:
\begin{align}
u_t^S & =\sum_{j=-1}^{1}\sum_{n,k=-\infty}^{\infty}u_{t,jnk}^Se^{-i\left(j \Upsilon_{s}+n \Upsilon_{r}+k \Upsilon_{\theta}\right)\lambda}\;,\label{eq:uts}
\\ u_{\phi}^S  &=\sum_{j=-1}^{1}\sum_{n,k=-\infty}^{\infty}u_{\phi,jnk}^Se^{-i\left(j \Upsilon_{s}+n \Upsilon_{r}+k \Upsilon_{\theta}\right)\lambda}\;.\label{eq:uphis}
\end{align}
We split $u_t^S$ into a constant $u_{t,0}^S$ plus an oscillatory contribution $\delta u_{t}^S(\lambda)$:
\begin{align}
u_t^S & =u_{t,0}^S+\delta u_{t}^S(\lambda)\;.\ \ \ 
\label{eq:deltauts}
\end{align}
We divide $u_{\phi}^S$ in the same way:
\begin{align}u_{\phi}^S =u_{\phi, 0}^S+\delta u_{\phi}^S(\lambda)\;\label{eq:deltauphis}
\end{align}
Using Eqs.\ (\ref{eq:Reqs}), we can immediately solve for $\delta u^S_t$ and $\delta u^S_{\phi}$.

We also use a frequency-domain description to solve the radial and polar Eqs.\ (\ref{eq:forcer}) -- (\ref{eq:forcetheta}). As described in Sec.\ \ref{sec:spinningbodydescrip}, generic orbits can be parameterized by
\begin{align}
    r & =\frac{p M}{1 + e\cos\chi_r} + \delta\mathcalligra{r}_S\;,\label{eq:rgenparam1} \\
    \cos\theta &= \sin I\cos\chi_\theta + \delta\mathcalligra{z}_S\;.\label{eq:thetagenparam1}
\end{align}
We break the radial true anomaly $\chi_r$ in Eq.\ (\ref{eq:rgenparam1}) into a mean anomaly $w_r=\Upsilon_r\lambda$ and oscillating contributions $\delta\chi_{r}$; we break up the the polar true anomaly $\chi_\theta$ in Eq.\ (\ref{eq:thetagenparam1}) similarly, using $w_\theta=\Upsilon_\theta\lambda$:
 \begin{align}
 \chi_r&= w_r+\delta\chi_r\;, \ \ \chi_\theta= w_\theta+\delta\chi_\theta\;.
 \end{align}
The mean anomalies have geodesic and spin-curvature pieces,
\begin{align}
w_{r} & =\left(\hat{\Upsilon}_r + \Upsilon^S_r\right)\lambda\;,\ \ w_{\theta} =\left(\hat{\Upsilon}_\theta + \Upsilon^S_\theta\right)\lambda\;,\label{eq:meananomtheta}
\end{align}
where $\Upsilon^S_r$ is the contribution to the radial Mino-time frequency arising from spin-curvature coupling, and  $\Upsilon^S_\theta$ is the analogous contribution to the polar Mino-time frequency.  The oscillating contributions likewise have one piece that arises from geodesic motion $\delta \hat{\chi}_x$ and another associated with spin-curvature coupling $\delta \chi^S_x$, where $x\in[r,\theta]$:
\begin{align}
\delta\chi_{r}  =\delta \hat{\chi}_r(w_r) + \delta \chi^S_r\;, \ \ \delta\chi_{\theta}  =\delta \hat{\chi}_\theta(w_\theta) + \delta \chi^S_\theta\;.\label{eq:deltachi}
\end{align}
In Eq.\ (\ref{eq:deltachi}), the Fourier coefficients of $\delta \hat{\chi}_r(w_r)$ and $\delta \hat{\chi}_\theta(w_\theta)$ are identical to those used to describe the anomaly angle of a geodesic orbit with parameters $p$, $e$ and $I$ in Eqs.\ (\ref{eq:rdef}) and  (\ref{eq:thdef}): 
\begin{align}
\delta \hat{\chi}_{r}(w_r) &=\sum_{n=-\infty}^{\infty}\delta \hat{\chi}_{r,n} e^{-in w_r}\;,\label{eq:deltachihatr} \\
\delta \hat{\chi}_{\theta}(w_\theta) &=\sum_{k=-\infty}^{\infty}\delta \hat{\chi}_{\theta,k} e^{-ik w_\theta}\;. \label{eq:deltachihattheta}
\end{align}
Note, however, that the phases $w_r$ and $w_\theta$ are not the same as those for the geodesic orbit with corresponding values of $(p,e,I)$, due to the presence of $\Upsilon^S_r$ and $\Upsilon^S_\theta$ in Eq.\ (\ref{eq:meananomtheta}).  The spin-corrections to the fundamental frequencies are built into our parameterization of spinning-body orbits. We explicitly include the $w_r$ and $w_\theta$ arguments in Eq.\ (\ref{eq:deltachi}) to emphasize this.

\subsubsection{Reference geodesics}
\label{sec:RefGeodFreqDom}

As we have discussed, we cannot in general constrain the radial or polar motion of spinning body orbits to lie between two fixed turning points as we can for bound geodesics.  However, we can constrain the purely radial motion (aspects of the motion that only involve harmonics of $\Upsilon_{r}$) and the purely polar motion (with only harmonics in $\Upsilon_{\theta}$) to lie within the radial and polar turning points of a given geodesic orbit.  In our approach, we parameterize an orbit by selecting a geodesic with parameters $(p,e,I)$, as well as an initial spin-vector orientation.  The purely radial motion of the spinning body's motion is then confined to the region $pM/(1 + e) \le r \le pM/(1 - e)$, and its purely polar is confined to $-\sin{I} \le \cos\theta \le \sin{I}$.  We call the geodesic with parameters $(p,e,I)$ in this picture the ``reference geodesic.''  We briefly introduced this concept in Sec.\ \ref{sec:spinningbodydescrip}.  Note that there are alternative mappings between geodesics and spinning bodies that have been used in the literature; see Appendix \ref{sec:referencegeodesics} for further discussion. 

We write $\delta \chi^S_r$ and $\delta \chi^S_\theta$ as Fourier expansions,
\begin{align}
\delta \chi^S_{r} &=\sum_{n=-\infty}^{\infty}\delta \chi^S_{r,n} e^{-in w_r}\;,
\label{eq:deltachiSr}\\
\delta \chi^S_{\theta} &=\sum_{k=-\infty}^{\infty}\delta \chi^S_{\theta,k}e^{-ik w_\theta}\;. \label{eq:deltachiStheta}
\end{align}
Note that because $\delta\chi^S_r$ and $\delta\chi^S_\theta$ both have average values of zero (they represent oscillatory contributions to the $\chi^S_r$ and $\chi^S_\theta$, we set $\delta\chi^S_{r,0} = 0$ and $\delta\chi^S_{\theta,0} = 0$).  Notice that the expansion for $\delta \chi^S_{r}$ in Eq.\ (\ref{eq:deltachiSr}) consists purely harmonics at the radial frequency; $\delta \chi^S_{\theta}$ in Eq.\ (\ref{eq:deltachiStheta}) likewise consists purely of harmonics at the polar frequency. In this way, we have constrained the purely radial motion to the interval $p/(1+e)\leq r \leq p/(1-e)$ and purely polar motion to the interval $-\sin I\leq\cos\theta\leq\sin I $. 

The remaining dynamics, consisting of motion that is neither purely radial nor purely polar, describes how the libration regions varies, and is mapped onto the quantities $\delta \mathcalligra{r}_S$ and $\delta \mathcalligra{z}_S$.  We expand these quantities using generic Fourier expansions of the form shown in Eq.\ (\ref{freqexpand}):
\begin{align}
\delta\mathcalligra{r}_S &=\sum_{j=-1}^{1}\sum_{n,k = -\infty}^{\infty} \delta\mathcalligra{r}_{S,jnk} e^{-i\left(jw_{s} + nw_{r} + kw_{\theta}\right)}\;, \label{eq:deltachicircinclprecessr20} \\
\delta\mathcalligra{z}_S &=\sum_{j=-1}^{1}\sum_{n,k = -\infty}^{\infty} \delta\mathcalligra{z}_{S,jnk} e^{-i\left(jw_{s} + nw_{r} + kw_{\theta}\right)}\;,\label{eq:deltachicircinclprecesstheta20} 
\end{align}
where $w_s=\Upsilon_s\lambda$.  Notice that harmonics of all three frequencies --- radial, polar, and spin precession --- are present in these expansions.  When we evaluate Eq.\ (\ref{eq:deltachicircinclprecessr20}), we require that $k$ and $j$ cannot both be zero; otherwise, that contribution would represent a purely radial dynamic, which we have constrained to be in the anomaly angle $\chi_r$.  Likewise, when we evaluate Eq.\ (\ref{eq:deltachicircinclprecesstheta20}), we require that $n$ and $j$ cannot both be zero, since the purely polar dynamics is entirely contained in $\chi_\theta$. 

In summary, the anomaly angles $\delta\chi_r^S$ and $\delta\chi_{\theta}^S$ control the shape of the orbit while keeping the turning points unchanged relative to the reference geodesic orbit, whereas $\delta\mathcalligra{r}_S$ and $\delta\mathcalligra{z}_S$ affect the position of the turning points and introduce spin precession effects into the dynamics.  In the nearly equatorial case ($I=0$), we find $\delta\mathcalligra{z}_S= -\delta\theta_S$; in the nearly circular case ($e=0$), we have $\delta\mathcalligra{r}_S = \delta r_S$.

\subsubsection{Deviation of a spinning body's orbit from its reference geodesic}

Once we expand the anomaly variables $\chi_r$ and $\chi_\theta$ as discussed in Sec.\ \ref{sec:generalities}, a generic orbit's radial and polar motion as described by Eqs.\ (\ref{eq:rgenparam1}) and (\ref{eq:thetagenparam1}) can be written in the form
\begin{align}
r(\lambda) &= \frac{pM}{1+e\cos(w_r+\delta \hat{\chi}_r(w_r)+\delta \chi^S_r)} + \delta\mathcalligra{r}_S\;,
\label{rparamagain}
\\
\cos\theta(\lambda) &= \sin(I)\cos(w_\theta+\delta \hat{\chi}_\theta(w_\theta)+\delta \chi^S_\theta) + \delta\mathcalligra{z}_S\;.
\label{thparamagain}
\end{align}
Here, $\delta \hat{\chi}_r(w_r)$, $\delta \chi^S_r$  and $\delta\mathcalligra{r}_S$ are given by Eqs.\ (\ref{eq:deltachihatr}), (\ref{eq:deltachiSr}) and (\ref{eq:deltachicircinclprecessr20}); the analogous quantities for the polar motion $\delta\hat\chi_\theta(w_\theta)$, $\delta\chi_\theta^S$, and $\delta\mathcalligra{z}_S$ are given by Eqs.\ (\ref{eq:deltachihattheta}), (\ref{eq:deltachiStheta}) and (\ref{eq:deltachicircinclprecesstheta20}).  Notice that the functions $\delta \hat{\chi}_r(w_r)$ and $\delta \hat{\chi}_\theta(w_\theta)$ have as their arguments $w_r$ and $w_\theta$, whose forms are given in Eq.\ (\ref{eq:meananomtheta}).  These functions are exactly the oscillating contributions to the anomaly angles that one computes for geodesic orbits, but with their frequencies shifted to remain phase-locked with spinning-body orbits.  For geodesics, their arguments would be $\hat w_r = \hat\Upsilon_r\lambda$ and $\hat w_\theta = \hat\Upsilon_\theta\lambda$.

In Sec.\ \ref{sec:SpinDev}, we defined the deviation from the geodesic trajectory induced by the spin of the small body by writing $x^\alpha(\lambda)$ as
\begin{equation}
\delta x^\alpha_S(\lambda) = x^\alpha(\lambda) - \hat{x}^\alpha(\lambda)\;, 
\label{eq:deltaxSdefinition}
\end{equation}
where $\hat{x}^\alpha(\lambda)$ is a geodesic orbit.  Using the reference geodesic in this equation, $\hat r$ and $\hat \theta$ are given by 
\begin{align}
    \hat{r}(\lambda)=\frac{pM}{1+e\cos \left(\hat{w}_r+\delta \hat{\chi}_r(\hat{w}_r)\right)}\;,
    \label{eq:refgeodradial}\\
    \cos \hat{\theta}(\lambda)=\sin I\cos\left(\hat{w}_\theta+\delta \hat{\chi}_\theta(\hat{w}_\theta)\right)\;.
    \label{eq:refgeodpolar}
\end{align}
Here we use the purely geodesic forms
\begin{align}
\delta \hat{\chi}_{r}(\hat{w}_r) &=\sum_{n=-\infty}^{\infty}\delta \hat{\chi}_{r,n} e^{-in \hat{w}_r\lambda}\;,\label{eq:deltachihatr2} \\
\delta \hat{\chi}_{\theta}(\hat{w}_\theta) &=\sum_{k=-\infty}^{\infty}\delta \hat{\chi}_{\theta,k} e^{-ik \hat{w}_\theta\lambda}\;. \label{eq:deltachihattheta2}
\end{align}
Combining the definition (\ref{eq:deltaxSdefinition}) with our solutions for the spinning body's motion, Eqs.\ (\ref{rparamagain}) and (\ref{thparamagain}), and for the reference geodesic, Eqs.\ (\ref{eq:refgeodradial}) and (\ref{eq:refgeodpolar}), we find
\begin{align}
    \delta r_S(\lambda) &= epM\frac{\Upsilon_r^S\lambda\left(1 - i\sum_{n}n\delta \hat{\chi}_{r,n} e^{-in\hat{w}_r}\right) + \delta \chi^S_r(w_r)}{\left[1 + e\cos\left(\hat{w}_r + \delta \hat{\chi}_r(\hat{w}_r)\right)\right]^2}
    \nonumber\\
    &\times \sin\left(\hat{w}_r + \delta\hat{\chi}_r(\hat{w}_r)\right) + \delta\mathcalligra{r}_S\;,  \label{eq:secularr}\\
    \delta \theta_S(\lambda) &= \Upsilon_r^S\lambda\left(1 - i\sum_{k}k\delta \hat{\chi}_{\theta,k} e^{-ik\hat{w}_\theta}\right) + \delta \chi_\theta^S(w_\theta)
    \nonumber\\
    &-\frac{\delta\mathcalligra{z}_S}{\sin I\sin\left(\hat w_\theta + \delta \hat\chi_\theta(\hat w_\theta)\right)}\;.
    \label{eq:seculartheta}
\end{align}
Notice that both $\delta r_S(\lambda)$ and $\delta \theta_S(\lambda)$ show secular growth.  This is because of the difference in frequencies between the geodesic $\hat{x}^\alpha(\lambda)$ and spinning-body $x^\alpha(\lambda)$ orbits.  The presence of these secularly growing terms means that, as defined, $\delta r_S(\lambda)$ and $\delta \theta_S(\lambda)$ cannot easily be studied using a frequency-domain treatment.

To address this, consider a slightly modified version of this definition:
\begin{equation}
    \delta x^\alpha_{S,{\rm shift}}(\lambda) = x^\alpha(\lambda)- \hat x^\alpha_{\rm shift}(\lambda)\;.
\end{equation}
This deviation is defined versus a {\it frequency-shifted} formulation of the geodesic motion:
\begin{align}
    \hat{r}_{\rm shift}(\lambda)=\frac{pM}{1+e\cos \left(w_r+\delta \hat{\chi}_r(w_r)\right)}\;,
    \label{eq:refgeodradial_shift}\\
    \cos \hat{\theta}_{\rm shift}(\lambda)=\sin I\cos\left(w_\theta+\delta \hat{\chi}_\theta(w_\theta)\right)\;.
    \label{eq:refgeodpolar_shift}
\end{align}
Equations (\ref{eq:refgeodradial_shift}) and (\ref{eq:refgeodpolar_shift}) describe a trajectory that is identical to the reference geodesic, but with all periodic features oscillating at the frequency associated with the spinning body's orbit.  The deviation from this shifted geodesic is given by
\begin{align}
    \delta r_{S, {\rm shift}}(\lambda) &= epM\frac{\delta \chi_r^S(w_r)\sin\left[w_r + \delta\hat{\chi}_r(w_r)\right]}{\left(1 + e\cos\left[w_r + \delta\hat{\chi}_r(w_r)\right]\right)^2} + \delta\mathcalligra{r}_S\;,
    \label{eq:deltarSshift}\\
    \delta\theta_{S,{\rm shift}} &= \delta \chi_\theta^S(w_\theta) - \frac{\delta\mathcalligra{z}_S}{\sin I\sin\left(\hat w_\theta + \delta \hat\chi_\theta(\hat w_\theta)\right)}\;.
    \label{eq:deltathSshift}
\end{align}
We discuss a variant of Eq.\ (\ref{eq:deltarSshift}) which does not include the libration shift $\delta\mathcalligra{r}_S$ in Appendix A of our companion paper \cite{Paper1}.  These modified offsets from the reference geodesic do not exhibit any secular growth, and can be nicely described using this paper's frequency-domain expansions.

\subsubsection{Coordinate-time quantities}
\label{sec:coordinatetime}

We can use our spinning-body solutions for $u_{\phi}$ to compute the Mino-time $\phi$-frequency $\Upsilon_{\phi}$, using 
\begin{equation}
\Upsilon_\phi^S=U_{S,000}^{\phi}\;,
\end{equation}
where 
\begin{equation}
U_{S,000}^{\phi}= \frac{1}{\Lambda_r \Lambda_{\theta} \Lambda_{s}} \int_0^{\Lambda_r}\int_0^{\Lambda_{\theta}}\int_0^{\Lambda_s} U_{S}^{\phi} \ d \lambda_r d \lambda_{\theta} d \lambda_{s}\;;
\end{equation}
we remind the reader that $U^\phi \equiv d\phi/d\lambda$.  Similarly, we can calculate the spin-correction to $\Gamma$ which denotes the average rate of accumulation of coordinate-time $t$ per unit Mino-time using
\begin{equation}
\Gamma^S=U_{S,000}^{t}\;,
\end{equation}
where 
\begin{equation}
U_{S,000}^{t}= \frac{1}{\Lambda_r \Lambda_{\theta} \Lambda_{s}} \int_0^{\Lambda_r}\int_0^{\Lambda_{\theta}}\int_0^{\Lambda_s} U_{S}^{t} \ d \lambda_r d \lambda_{\theta} d \lambda_{s}\;.
\end{equation}
Once we have the correction to $\Gamma$, we can convert any of the Mino-time frequencies into coordinate-time frequencies. Observing that
\begin{equation}
\hat\Omega_k + \Omega^S_k = \frac{\hat\Upsilon_k + \Upsilon_k^S}{\hat\Gamma + \Gamma^S}
\end{equation}
for $k\in(r,\theta,\phi)$, we see that shifts to the coordinate-time frequencies are given by 
\begin{equation}
\Omega_k^S = \hat \Omega_k\left(\frac{\Upsilon_k^S}{\hat\Upsilon_k}-\frac{\Gamma^S}{\hat\Gamma}\right)
\end{equation}
to linear order in the small body's spin.

\subsection{Results}
\label{sec:results}

\subsubsection{Nearly circular orbits: Aligned spin}
\label{sec:circinclalign}

We now discuss spinning-body orbits that are $\mathcal{O}(S)$ away from being circular --- nearly circular orbits. We outline how we compute the first-order in spin contribution to the polar Mino-time frequency $\Upsilon^S_{\theta}$ using a frequency-domain description for the motion. 

We consider a circular inclined reference geodesic, with the spin vector of the small body aligned with the orbit. In this case, orbits can be described using expansions of the form 
\begin{align}
f(\lambda) & =\sum_{k=-\infty}^{\infty}f_k e^{-ik w_\theta} \label{eq:polarexp}\;.
\end{align}In order to evaluate these expressions, we truncate the Fourier expansion at a finite value; for the expansion above, we truncate the series at $k_{\text{max}}$. By truncating this Fourier expansion at an appropriately large $k_{\text{max}}$, we can compute orbits with an arbitrarily high inclination. 

\begin{figure*}
\centerline{\includegraphics[scale=0.53]{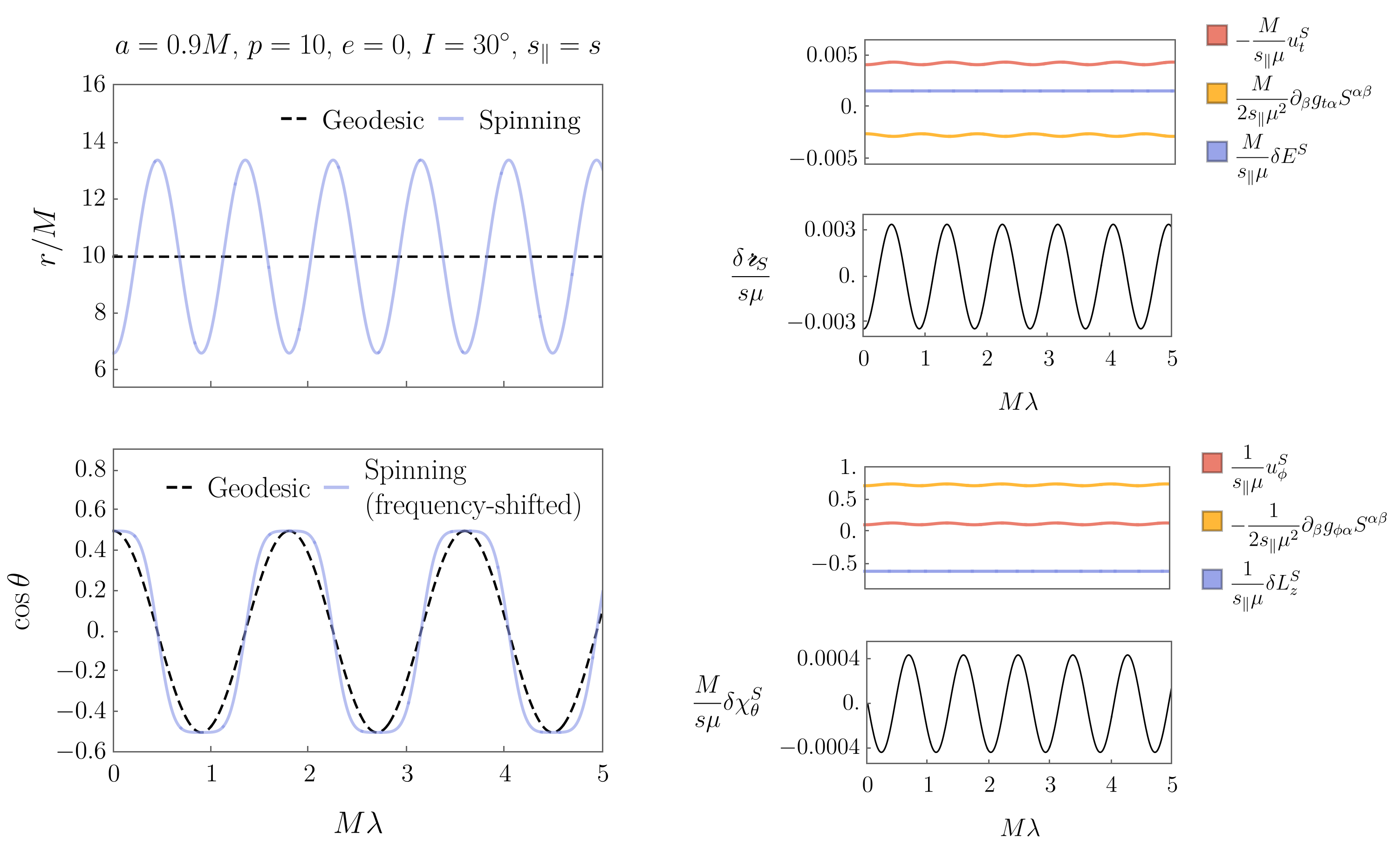}}
\caption{Example of the motion of a nearly circular orbit for an aligned spinning test body around a Kerr black hole with $a=0.9M$. Top left panel shows $r$ versus $\lambda$ for a geodesic (black dashed) and a spinning test body (blue solid) orbit. The radial reference geodesic is circular, with $p=10$, $e=0$. Bottom left panel shows $\cos\theta$ versus $\lambda$ for a geodesic (black dashed) and a spinning test body (blue solid) orbit. These orbits share polar turning points, corresponding to $I=30^{\circ}$.  Note that, in the left two panels, we have used an unphysically high spin $\mu s /M=10^3$ in order make the spin-curvature effects clearly visible.  Also note that the spinning-body orbit has been shifted slightly: its polar frequency $\Upsilon_\theta= \hat\Upsilon_\theta + \Upsilon_\theta^S$ has been replaced with $\hat\Upsilon_\theta$.  This is done so that the geodesic and the spinning-body orbit pass through their polar turning points at the same times, which helps to illustrate differences in their motion between each turning point. Top right shows $-u_t^S$ (red), $\partial_{\beta}g_{t\alpha}S^{\alpha\beta}/2\mu$ (orange), $\delta E^S$ (blue) as well as $\delta \mathcalligra{r}_S$ (black), all versus $\lambda$. Finally, the bottom right panel shows $u_{\phi}^S$ (red), $-\partial_{\beta}g_{\phi\alpha}S^{\alpha\beta}/2\mu$ (orange), $\delta L_z^S$ (blue) as well as $\delta\chi_{\theta}^S$ (black), all versus $\lambda$. Notice that the spin-induced shifts to the integrals of motion $E$ and $L_z$ are constants, although each such term has contributions that oscillate. In making these plots, we have used $s=s_\parallel$ and $k_{\text{max}}=6$.
\label{fig:exampleorbitscircincl}}
\end{figure*}

As described in Sec.\ \ref{sec:spinningbodydescrip}, we use a parameterization to describe the motion in $\theta$ which resembles the form typically used to describe geodesic orbits, as in Eq.\ (\ref{eq:thdef}).  In addition to this, we must account for the fact that the spin of the test body induces oscillations about $pM$, the radius of the circular reference geodesic.  We thus parameterize the orbit as
\begin{align}
r & =pM + \delta\mathcalligra{r}_S\;,
\label{eq:paramcircinclr}\\
\cos\theta & =\sin I\cos \left(w_{\theta}+\delta\hat{\chi}_{\theta}(w_\theta)+ \delta\chi_{\theta}^S\right)\;.\label{eq:paramcircincltheta}
 \end{align}
The functions $\delta\chi_{\theta}^{S}$ and $\delta\mathcalligra{r}_S$ are described by purely polar oscillations in this case:
\begin{align}
\delta\chi_{\theta}^{S} & =\sum_{k=-\infty}^{\infty}\delta\chi_{\theta,k}^{S}e^{-ik w_{\theta}}\;,\label{eq:deltachicircinclr} \\
\delta\mathcalligra{r}_S& =\sum_{k=-\infty}^{\infty}\delta\mathcalligra{r}_{S,k} e^{-ikw_{\theta}}\;.
\label{eq:deltachicircincltheta}
\end{align}
\begin{widetext}
We insert Eqs.\ (\ref{eq:paramcircinclr}), (\ref{eq:paramcircincltheta}) and (\ref{eq:utuphi}) into (\ref{eq:forcer}) -- (\ref{eq:forcetheta}) and linearize in spin.  The first-order-in-spin piece of Eq.\ (\ref{eq:forcer}) becomes
\begin{align}
\mathcal{F}_{\mathcalligra{r}}\frac{d^2 \delta\mathcalligra{r}_S}{d \lambda^2}+\mathcal{G}_{\mathcalligra{r}}\frac{d \delta\mathcalligra{r}_S}{d \lambda}+\mathcal{G}_\theta  \frac{d \delta\chi_{\theta}^S}{d \lambda}+\mathcal{H}_{\mathcalligra{r}} \delta\mathcalligra{r}_S+\mathcal{H}_\theta\delta\chi_{\theta}^S+\mathcal{I}_{1\theta} \Upsilon_{\theta}^S +\mathcal{I}_2 u^S_{t,0} +\mathcal{I}_3 u^S_{\phi, 0}+\mathcal{J} =0\;,\label{eq:radiallinMP3}
\end{align}
where $\mathcal{F}_{\mathcalligra{r}}$, $\mathcal{G}_{\mathcalligra{r}}$, $\mathcal{G}_\theta$, $\mathcal{H}_{\mathcalligra{r}}$,  $\mathcal{H}_\theta$, $\mathcal{I}_{1\theta}$, $\mathcal{I}_{2}$, $\mathcal{I}_{3}$ and  $\mathcal{J}$ are all functions of known quantities evaluated on geodesics.  We now consider the first-order-in-spin piece of Eq.\ (\ref{eq:forcetheta}), which becomes
\begin{align}
\mathcal{Q}_\theta\frac{d^2 \delta\chi_{\theta}^S}{d \lambda^2}+\mathcal{S}_{\mathcalligra{r}} \frac{d \delta\mathcalligra{r}_S}{d \lambda}+\mathcal{S}_\theta\frac{d \delta\chi_{\theta}^S}{d \lambda}+\mathcal{T}_{\mathcalligra{r}}\delta\mathcalligra{r}_S+\mathcal{T}_{\theta}\delta\chi_{\theta}^S+\mathcal{U}_{1\theta}\Upsilon_{\theta}^S +\mathcal{U}_{2}u^S_{t,0} +\mathcal{U}_{3}u^S_{\phi, 0}+\mathcal{V}=0\;,\label{eq:polarlinMP3}
\end{align}where $\mathcal{Q}_\theta$, $\mathcal{S}_{\mathcalligra{r}}$, $\mathcal{S}_\theta$, $\mathcal{T}_{\mathcalligra{r}}$,  $\mathcal{T}_{\theta}$, $\mathcal{U}_{1\theta}$, $\mathcal{U}_{2}$, $\mathcal{U}_{3}$ and $\mathcal{V}$ are all functions of known quantities on geodesics. Third, we use the constraint $u^{\alpha}u_{\alpha}=-1$ to obtain a linearized equation of the form
\begin{align}
\mathcal{K}_{\mathcalligra{r}} \frac{d \delta\mathcalligra{r}_S}{d \lambda}+\mathcal{K}_\theta \frac{d \delta\chi_{\theta }^S}{d \lambda}+\mathcal{M}_{\mathcalligra{r}} \delta\mathcalligra{r}_S+\mathcal{M}_\theta \delta\chi_{\theta }^S+\mathcal{N}_{1\theta}\Upsilon_{\theta}^S +\mathcal{N}_{2}u^S_{t,0} +\mathcal{N}_{3} u^S_{\phi, 0}+\mathcal{P}=0\;,\label{eq:udotu3}
\end{align}
where $\mathcal{K}_{\mathcalligra{r}}$,  $\mathcal{K}_\theta$, $\mathcal{M}_{\mathcalligra{r}}$, $\mathcal{M}_\theta$, $\mathcal{N}_{1\theta}$, $\mathcal{N}_{2}$, $\mathcal{N}_{3}$ and $\mathcal{P}$ are again all functions\footnote{The functions $\mathcal{F}_{\mathcalligra{r}}$, $\mathcal{G}_{\mathcalligra{r}}$, etc.\ follow a mostly alphabetic sequence; however, we skip the letter $\mathcal{L}$ in our scheme to avoid confusion with the angular momentum 4-vector defined in Eq.\ (\ref{eq:orbangmomdef}).} of known quantities on geodesics. 
\end{widetext}

We write the functions $\mathcal{F}_{\mathcalligra{r}}$, $\mathcal{G}_{\mathcalligra{r}}$, $\mathcal{G}_\theta$, $\mathcal{H}_{\mathcalligra{r}}$, $\mathcal{H}_\theta$, $\mathcal{I}_{1\theta}$, $\mathcal{I}_{2}$, $\mathcal{I}_{3}$, $\mathcal{J}$, $\mathcal{Q}_\theta$, $\mathcal{S}_{\mathcalligra{r}}$, $\mathcal{S}_\theta$, $\mathcal{T}_{\mathcalligra{r}}$, $\mathcal{T}_{\theta}$, $\mathcal{U}_{1\theta}$, $\mathcal{U}_{2}$, $\mathcal{U}_{3}$, $\mathcal{V}$,  $\mathcal{K}_{\mathcalligra{r}}$, $\mathcal{K}_\theta$, $\mathcal{M}_{\mathcalligra{r}}$, $\mathcal{M}_\theta$, $\mathcal{N}_{1\theta}$, $\mathcal{N}_{2}$, $\mathcal{N}_{3}$ and $\mathcal{P}$ as Fourier expansions of the form (\ref{eq:polarexp}). The explicit forms for many of these expressions in the limiting case of nearly equatorial Schwarzschild orbits ($a=0$) can be found in Appendix C of Ref.\ \cite{Paper1}. For the general case, we provide expressions using the \textit{Mathematica} notebook in this paper's Supplemental Material \cite{SupplementalMaterial}.  Some of the general expressions are very lengthy (hundreds of terms long) and could likely be simplified with some effort; we present them in a companion Mathematica notebook for convenience and completeness.

We insert these expansions along with (\ref{eq:deltachicircinclr}) and (\ref{eq:deltachicircincltheta}) into  (\ref{eq:radiallinMP3}), (\ref{eq:polarlinMP3}) and (\ref{eq:udotu3}). We then solve for the unknown variables $\delta\mathcalligra{r}_S$, $\delta\chi_{\theta}^S$, $\Upsilon_{\theta}^S$ $u_{t}^S$ and $u_{\phi}^S$.

In the left-hand panels of Fig.\ \ref{fig:exampleorbitscircincl}, we show $r$ and $\theta$ for a circular, inclined, spin-aligned orbit; $r$ and $\theta$ for the corresponding reference geodesic orbit are overplotted. The period associated with the spinning-body orbit's polar motion is shifted so that it remains phase-locked with the geodesic orbit.  The right-hand panels of Fig.\ \ref{fig:exampleorbitscircincl} show $\delta\mathcalligra{r}_S$ and $\delta\chi_\theta^S$ for the spinning-body orbit.  We also plot $u_t^S$ and $u_{\phi}^S$ alongside the spin contributions to the orbit's energy and axial angular momentum, $\delta E^S$ and $\delta L_z^S$.  Notice that the spinning-body orbit we obtain is not circular; this can be seen in the top left panel of Fig.\ \ref{fig:exampleorbitscircincl}, where the effect is exaggerated so that the oscillations in $r$ are clearly visible. In our companion paper Ref.\ \cite{Paper1}, we perturbed about an equatorial reference geodesic and obtained a spinning-body orbit that did not lie in the equatorial plane; here we perturb about a circular reference geodesic, yielding a corresponding spinning-body orbit that is not circular.  In contrast to the behavior we saw in Ref.\ \cite{Paper1}, we cannot attribute this behavior only to spin precession, since we see this effect even when the spin vector is aligned.

\begin{figure}
\centerline{\includegraphics[scale=0.52]{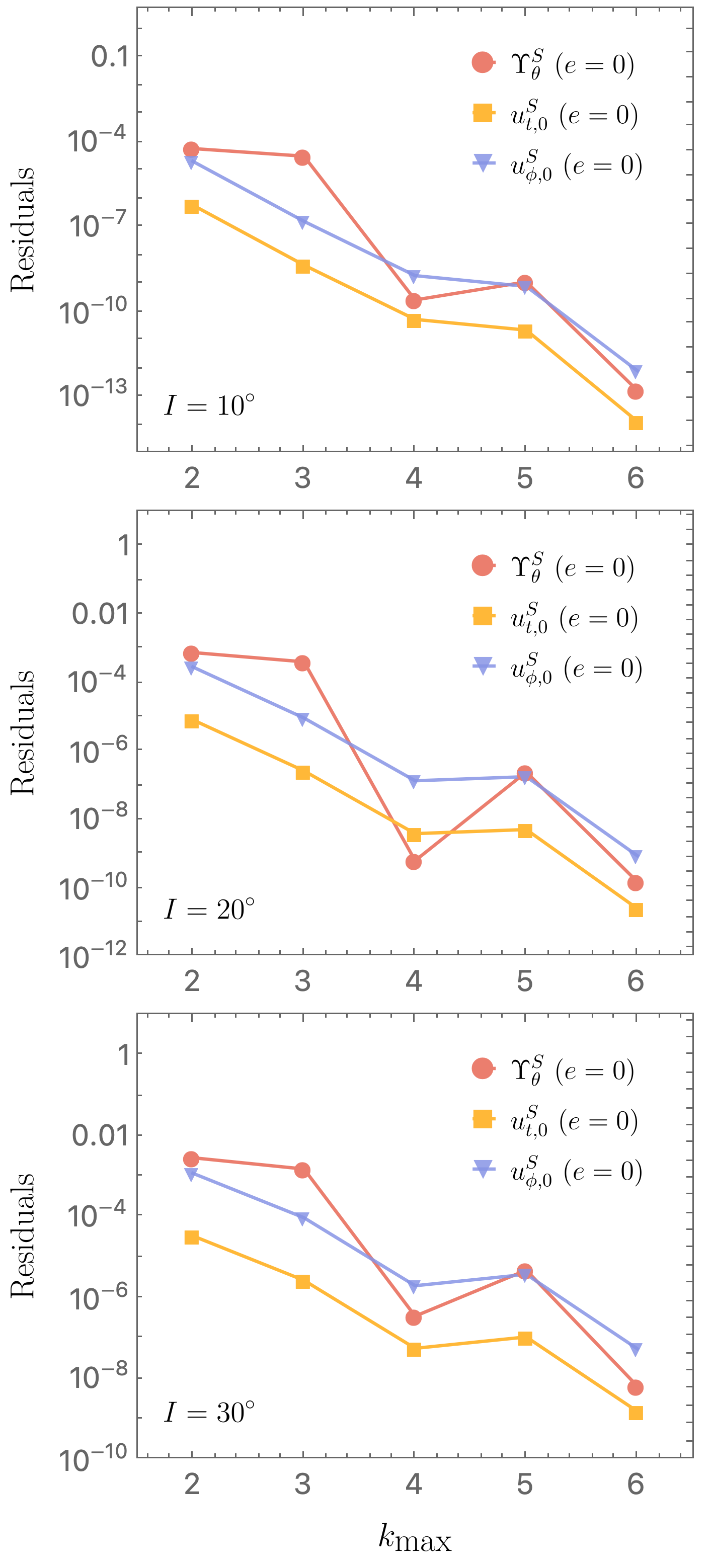}}
\caption{Plot of residuals versus $k_{\text{max}}$ for a nearly circular ($e=0$) orbit of an aligned ($s_\parallel=s$) spinning body. We plot $\Upsilon_{\theta}^S$, $u^S_{t,0}$, $u^S_{\phi,0}$ using red circular, orange square and blue triangular markers respectively. To compute these residuals, we evaluate the change between subsequent values of $k_\text{max}$ for each of the quantities plotted. Top panel shows $I=10^{\circ}$; middle shows $I=20^{\circ}$; and bottom shows $I=30^{\circ}$. In all cases, the large black hole has spin parameter $a=0.9M$, and the orbit has $p=10$. \label{fig:residualplotInearlycirc}}
\end{figure}

In Fig.\ \ref{fig:residualplotInearlycirc}, we see how $\Upsilon_{\theta}$, $u^S_{t,0}$ and $u^S_{\phi,0}$ converge to their true values in the case of nearly circular, inclined orbits. We define ``residuals" here to mean the difference between the value of the quantity computed at successive $k_{\text{max}}$'s, rather than a direct comparison with an exact value (as they were defined in Ref.\ \cite{Paper1}). As expected, the residuals generally decrease as $k_{\text{max}}$ increases. However, the pattern of convergence isn't strictly monotonic; the residuals tend to tick upwards for odd values  of $k_{\text{max}}$. 

\subsubsection{Nearly circular orbits: Misaligned spin}
\label{sec:circinclmisalign}

\begin{figure*}
\centerline{\includegraphics[scale=0.51]{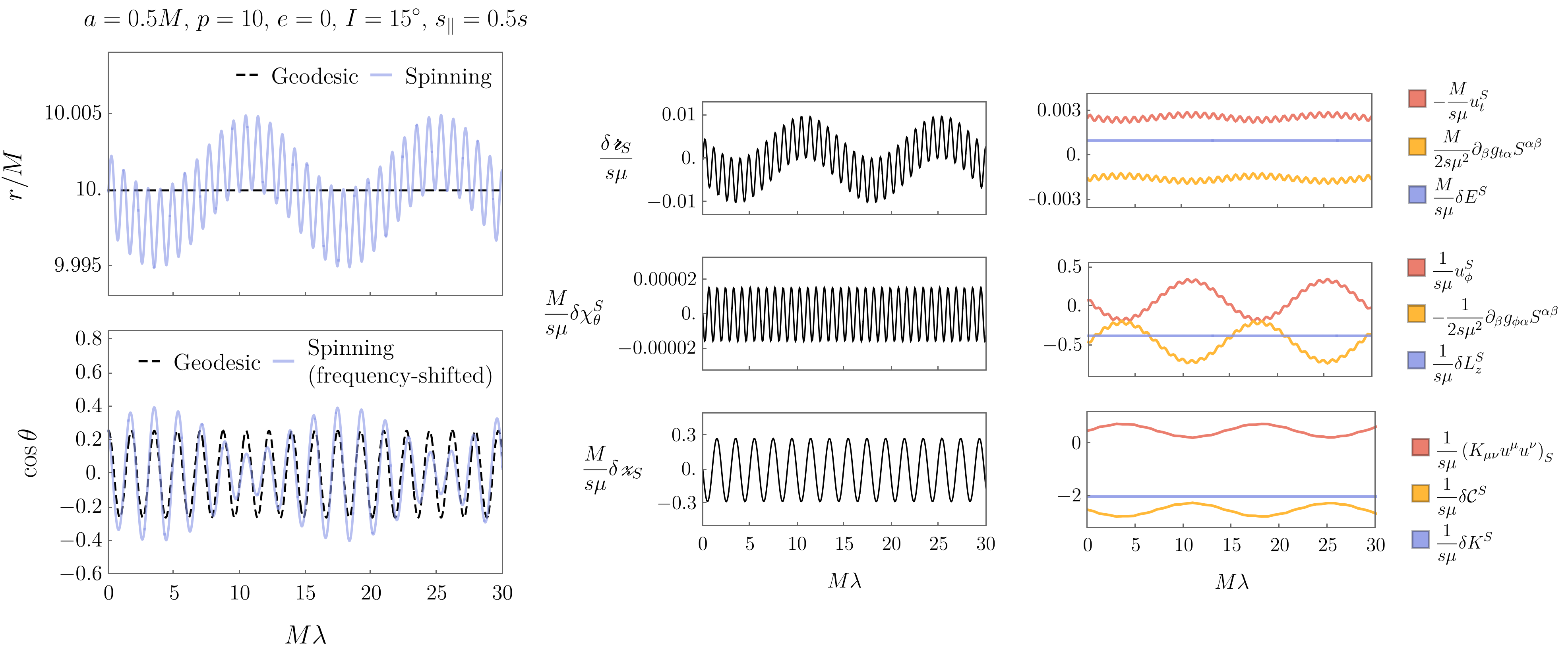}}
\caption{Example of the motion of a nearly circular orbit for a non-aligned spinning test body around a Kerr black hole with $a=0.5M$. Top left panel shows $r$ versus $\lambda$ for a geodesic (black dashed) and a spinning test body (blue solid) orbit. The radial reference geodesic is circular, with $p=10$, $e=0$. Bottom left panel shows $\cos\theta$ versus $\lambda$ for a geodesic (black dashed) and a spinning test body (blue solid) orbit. The polar reference geodesic has $I=15^{\circ}$. Note that, in the two left panels, we have used an unphysically high spin $\mu s /M=0.5$ in order make the spin-curvature effects clearly visible. Also note that for making this plot, the spinning-body orbit has been shifted slightly: its polar frequency $\Upsilon_\theta= \hat\Upsilon_\theta + \Upsilon_\theta^S$ has been replaced with $\hat\Upsilon_\theta $. This is done so that in the plot the geodesic and the spinning-body orbit pass through their polar turning points at the same times, which helps to illustrate differences in their motion between each turning point.  Middle column shows $\delta \mathcalligra{r}_S$, $\delta\chi_{\theta}^S$ and $\delta \mathcalligra{z}_S$, all versus $\lambda$ and all drawn with black solid lines. Top right panel shows $-u_t^S$ (red), $\partial_{\beta}g_{t\alpha}S^{\alpha\beta}/2\mu$ (orange), $\delta E^S$ (blue), all versus $\lambda$. Middle right panel shows $u_{\phi}^S$ (red), $-\partial_{\beta}g_{\phi\alpha}S^{\alpha\beta}/2\mu$ (orange), $\delta L_z^S$ (blue), all versus $\lambda$. Finally, bottom right panel shows $\left(K_{\mu\nu}u^{\mu}u^{\nu}\right)_S$ (red), $\delta\mathcal{C}^S$ (orange), $\delta K^S$ (blue), all versus $\lambda$. Notice that the spin-induced shifts to the integrals of motion $E$, $L_z$ and $K$ are constants, although each such term has contributions that oscillate. In making these plots, we have used $s_\parallel=s/2$, $s_\perp = \sqrt{3}s/2$, $\phi_s=\pi/2$ and $k_{\text{max}}=3$. \label{fig:exampleorbitscircinclprecess}}
\end{figure*}

We now consider nearly circular inclined orbits with the spin of the test body misaligned from the orbit (i.e., circular orbits $s_\perp \ne 0$).  Taking into account the effect of spin precession, many orbital quantities can be described using frequency-domain expansions of the form 
\begin{align}
f(\lambda) & =\sum_{j=-1}^{1}\sum_{k=-\infty}^{\infty}f_{jk} e^{-i\left(j w_s+k w_\theta\right)} \label{eq:polarexpprecess}\;.
\end{align}
As described in Sec.\ \ref{sec:spinningbodydescrip}, the parameterization of the orbit in this case has the form
\begin{align}
r & =pM+\delta\mathcalligra{r}_S\;,  \label{eq:paramcircinclprecessr}
\\ \cos\theta &=\sin I\cos \left(w_{\theta}+\delta\hat{\chi}_{\theta}(w_{\theta})+\delta\chi_{\theta}^S\right)+\delta\mathcalligra{z}_S\;.\label{eq:paramcircinclprecesstheta}
\end{align}
Compared to the parameterization in Sec.\ \ref{sec:circinclalign}, there is a new term $\delta\mathcalligra{z}_S$ which adjusts the polar turning points relative to the reference geodesic.  The libration variations $\delta\mathcalligra{r}_S$ and $\delta\mathcalligra{z}_S$ depend on both $\Upsilon_\theta$ and $\Upsilon_s$, while $\delta \chi^S_{\theta}$ only has oscillations at harmonics of $\Upsilon_\theta$:
\begin{align}
\delta\chi_{\theta}^{S} & =\sum_{k=-\infty}^{\infty}\delta\chi_{\theta,k}^{S}e^{-ik w_{\theta}}\;,\label{eq:deltachicircinclprecesstheta}  \\
\delta \mathcalligra{r}_S &=\sum_{j=-1}^{1}\sum_{k=-\infty}^{\infty}\delta \mathcalligra{r}_{S,jk}e^{-i\left(k w_{\theta}+j w_{s}\right)}\;,\label{eq:deltachicircinclprecessr2}  \\
\delta\mathcalligra{z}_S & =\sum_{j=-1}^{1}\sum_{k=-\infty}^{\infty}\delta\mathcalligra{z}_{S,jk} e^{-i\left(k w_{\theta}+j w_{s}\right)}\;,\label{eq:deltachicircinclprecesstheta2} 
\end{align}
where, in the last line, $j$ cannot equal zero. We then follow the same procedure as described for nearly circular inclined orbits with aligned spin to convert the time-domain expressions into a linear algebraic system in the frequency domain, but now including the term $\delta\mathcalligra{z}_S$ in the equations.  
\begin{widetext}
We insert equations (\ref{eq:paramcircinclprecessr}), (\ref{eq:paramcircinclprecesstheta}) and (\ref{eq:utuphi}) into (\ref{eq:forcer}) -- (\ref{eq:forcetheta}) and linearize in spin. Eq.\ (\ref{eq:forcer}) can be written
\begin{align}
\mathcal{F}_{\mathcalligra{r}} \frac{d^2 \delta\mathcalligra{r}_S}{d \lambda^2}+\mathcal{G}_{\mathcalligra{r}} \frac{d \delta\mathcalligra{r}_S}{d \lambda}+\mathcal{G}_\theta  \frac{d \delta\chi_{\theta}^S}{d \lambda}+\mathcal{G}_{\mathcalligra{z}}\frac{d \delta\mathcalligra{z}_S}{d \lambda}+\mathcal{H}_{\mathcalligra{r}} \delta\mathcalligra{r}_S
+\mathcal{H}_\theta\delta\chi_{\theta}^S+\mathcal{H}_{\mathcalligra{z}}\delta\mathcalligra{z}_S +\mathcal{I}_{1\theta} \Upsilon_{\theta}^S\nonumber \\ +\mathcal{I}_2 u^S_{t,0} +\mathcal{I}_3 u^S_{\phi, 0}+\mathcal{J} =0\;,\label{eq:radiallinMP4}
\end{align}
where $\mathcal{F}_{\mathcalligra{r}}$, $\mathcal{G}_{\mathcalligra{r}}$, $\mathcal{G}_\theta$, $\mathcal{G}_{\mathcalligra{z}}$,  $\mathcal{H}_{\mathcalligra{r}}$,  $\mathcal{H}_\theta$,  $\mathcal{H}_{\mathcalligra{z}}$, $\mathcal{I}_{1\theta}$, $\mathcal{I}_{2}$, $\mathcal{I}_{3}$ and $\mathcal{J}$ are all functions of known quantities evaluated on geodesics. Similarly, we can write Eq.\ (\ref{eq:forcetheta}) in the form 
\begin{align}
\mathcal{Q}_\theta\frac{d^2 \delta\chi_{\theta}^S}{d \lambda^2}+\mathcal{Q}_{\mathcalligra{z}}\frac{d^2 \delta\mathcalligra{z}_S}{d \lambda^2}+\mathcal{S}_{\mathcalligra{r}} \frac{d \delta\mathcalligra{r}_S}{d \lambda}+\mathcal{S}_\theta\frac{d \delta\chi_{\theta}^S}{d \lambda}+\mathcal{S}_{\mathcalligra{z}}\frac{d \delta\mathcalligra{z}_S}{d \lambda}+\mathcal{T}_{\mathcalligra{r}}\delta\mathcalligra{r}_S +\mathcal{T}_{\theta}\delta\chi_{\theta}^S
+\mathcal{T}_\mathcalligra{z}\delta\mathcalligra{z}_S \nonumber \\+\mathcal{U}_{1\theta}\Upsilon_{\theta}^S +\mathcal{U}_{2} u^S_{t,0} +\mathcal{U}_{3}u^S_{\phi, 0}+\mathcal{V}=0\;,\label{eq:polarlinMP4}
\end{align}
where $\mathcal{Q}_\theta$,  $\mathcal{Q}_{\mathcalligra{z}}$, $\mathcal{S}_{\mathcalligra{r}}$, $\mathcal{S}_\theta$, $\mathcal{S}_{\mathcalligra{z}}$, $\mathcal{T}_{\mathcalligra{r}}$,  $\mathcal{T}_{\theta}$, $\mathcal{T}_\mathcalligra{z}$, $\mathcal{U}_{1\theta}$, $\mathcal{U}_{2}$, $\mathcal{U}_{3}$ and $\mathcal{V}$ are all functions of known quantities evaluated on geodesics. We again also use $u^{\alpha}u_{\alpha}=-1$, yielding
\begin{align}
\mathcal{K}_{\mathcalligra{r}} \frac{d \delta\mathcalligra{r}_S}{d \lambda}+\mathcal{K}_\theta \frac{d \delta\chi_{\theta}^S}{d \lambda}+\mathcal{K}_{\mathcalligra{z}} \frac{d \delta\mathcalligra{z}_S}{d \lambda}+\mathcal{M}_{\mathcalligra{r}} \delta\mathcalligra{r}_S+\mathcal{M}_\theta \delta\chi_{\theta }^S+\mathcal{M}_{\mathcalligra{z}} \delta\mathcalligra{z}_S
+\mathcal{N}_{1\theta}\Upsilon_{\theta}^S+\mathcal{N}_{2}u^S_{t,0} +\mathcal{N}_{3} u^S_{\phi, 0}+\mathcal{P}=0\;,\label{eq:udotu4}
\end{align}
where $\mathcal{K}_{\mathcalligra{r}}$, $\mathcal{K}_\theta$, $\mathcal{K}_{\mathcalligra{z}}$, $\mathcal{M}_{\mathcalligra{r}}$, $\mathcal{M}_\theta$, $\mathcal{M}_{\mathcalligra{z}}$, $\mathcal{N}_{1\theta}$, $\mathcal{N}_{2}$, $\mathcal{N}_{3}$ and $\mathcal{P}$ are again all functions of known quantities evaluated on geodesics. 
\end{widetext}
We describe $\mathcal{F}_{\mathcalligra{r}}$, $\mathcal{G}_{\mathcalligra{r}}$, $\mathcal{G}_\theta$, $\mathcal{G}_{\mathcalligra{z}}$,  $\mathcal{H}_{\mathcalligra{r}}$, $\mathcal{H}_\theta$, $\mathcal{H}_{\mathcalligra{z}}$, $\mathcal{I}_{1\theta}$, $\mathcal{I}_{2}$, $\mathcal{I}_{3}$, $\mathcal{J}$, $\mathcal{Q}_\theta$, $\mathcal{Q}_{\mathcalligra{z}}$, $\mathcal{S}_{\mathcalligra{r}}$, $\mathcal{S}_\theta$, $\mathcal{S}_{\mathcalligra{z}}$, $\mathcal{T}_{\mathcalligra{r}}$, $\mathcal{T}_{\theta}$, $\mathcal{T}_\mathcalligra{z}$, $\mathcal{U}_{1\theta}$, $\mathcal{U}_{2}$, $\mathcal{U}_{3}$, $\mathcal{V}$, $\mathcal{K}_{\mathcalligra{r}}$, $\mathcal{K}_\theta$, $\mathcal{K}_{\mathcalligra{z}}$, $\mathcal{M}_{\mathcalligra{r}}$, $\mathcal{M}_\theta$, $\mathcal{M}_{\mathcalligra{z}}$, $\mathcal{N}_{1\theta}$, $\mathcal{N}_{2}$, $\mathcal{N}_{3}$ and $\mathcal{P}$ using Fourier expansions of the form (\ref{eq:polarexpprecess}).  We provide the full expressions for these functions in the \textit{Mathematica} notebook in the Supplemental Material accompanying this article \cite{SupplementalMaterial}.  We insert these expansions along with (\ref{eq:deltachicircinclprecesstheta}), (\ref{eq:deltachicircinclprecessr2}) and (\ref{eq:deltachicircinclprecesstheta2}), into (\ref{eq:radiallinMP4}), (\ref{eq:polarlinMP4}) and (\ref{eq:udotu4}). We then solve for the unknown variables $\delta\mathcalligra{r}_S$, $\delta\chi_{\theta}^S$, $\delta\mathcalligra{z}_S$,  $\Upsilon_{\theta}^S$, $u_{t}^S$ and  $u_{\phi}^S$. 

In the left-hand panels of Fig.\ \ref{fig:exampleorbitscircinclprecess}, we show $r$ and $\theta$ for a misaligned nearly circular spinning-body orbit, with the circular inclined reference geodesic overplotted for reference. As in Fig.\ \ref{fig:exampleorbitscircincl}, the spinning-body orbit's polar frequency is shifted so that it remains phase-locked with the geodesic orbit.  The form of $\delta\mathcalligra{r}_S$, $\delta\chi_{\theta}^{S}$ and $\delta\mathcalligra{z}_S$ for this orbit are shown in the right panels of Fig.\ \ref{fig:exampleorbitscircinclprecess}. As in Fig.\ \ref{fig:exampleorbitscircincl}, we  plot $u_t^S$ and $u_{\phi}^S$ as well as the corrections to the spinning body's orbital energy $\delta E^S$ and axial angular momentum $\delta L_z^S$ in the right panels of Fig.\ \ref{fig:exampleorbitscircinclprecess}. 

In the bottom right panel of Fig.\ \ref{fig:exampleorbitscircinclprecess}, we show the spin-correction to the Carter constant $K$. We plot the first-order in spin correction to the term $K_{\mu\nu}u^\mu u^\nu$ and the quantity $\delta \mathcal{C}^S$ which is defined in (\ref{eq:Cspin}), giving us the overall first-order correction to $K$ denoted $\delta K^S$.  For equatorial reference geodesics, $\delta Q^S$ has the simple form $2a s_{\parallel}$, as was discussed in our companion article Ref.\ \cite{Paper1}.  In this case, when the orbit is inclined and the spin vector is precessing, we find that the first-order in spin correction to $K_{\mu\nu}u^{\mu} u^{\nu}$ is no longer constant.  The oscillations in this quantity precisely cancel oscillations in $\delta \mathcal{C}^S$, yielding constant values for $\delta K^S$ and $\delta Q^S$.

\subsubsection{Generic orbits}
\label{sec:genericorbits}

We finally examine generic orbits of spinning test bodies.  We use the following Fourier expansion
\begin{align}
f(\lambda)=\sum_{n,k=-\infty}^{\infty}\sum_{j=-1}^1f_{jnk}e^{-i \left(j w_s+n w_r+k w_{\theta}\right)}\label{eq:Fexpandgeneric}
\end{align}
for the various quantities we must evaluate.  To evaluate these expressions, we truncate the Fourier expansion at a finite value; for the expansion above, we truncate the radial series at $n_{\text{max}}$ and the polar series at $k_{\text{max}}$. By truncating this Fourier expansion at an appropriately large $n_{\text{max}}$ and $k_{\text{max}}$, we can compute orbits with an arbitrarily high eccentricity and inclination. 

\begin{figure}
\centerline{\includegraphics[scale=0.52]{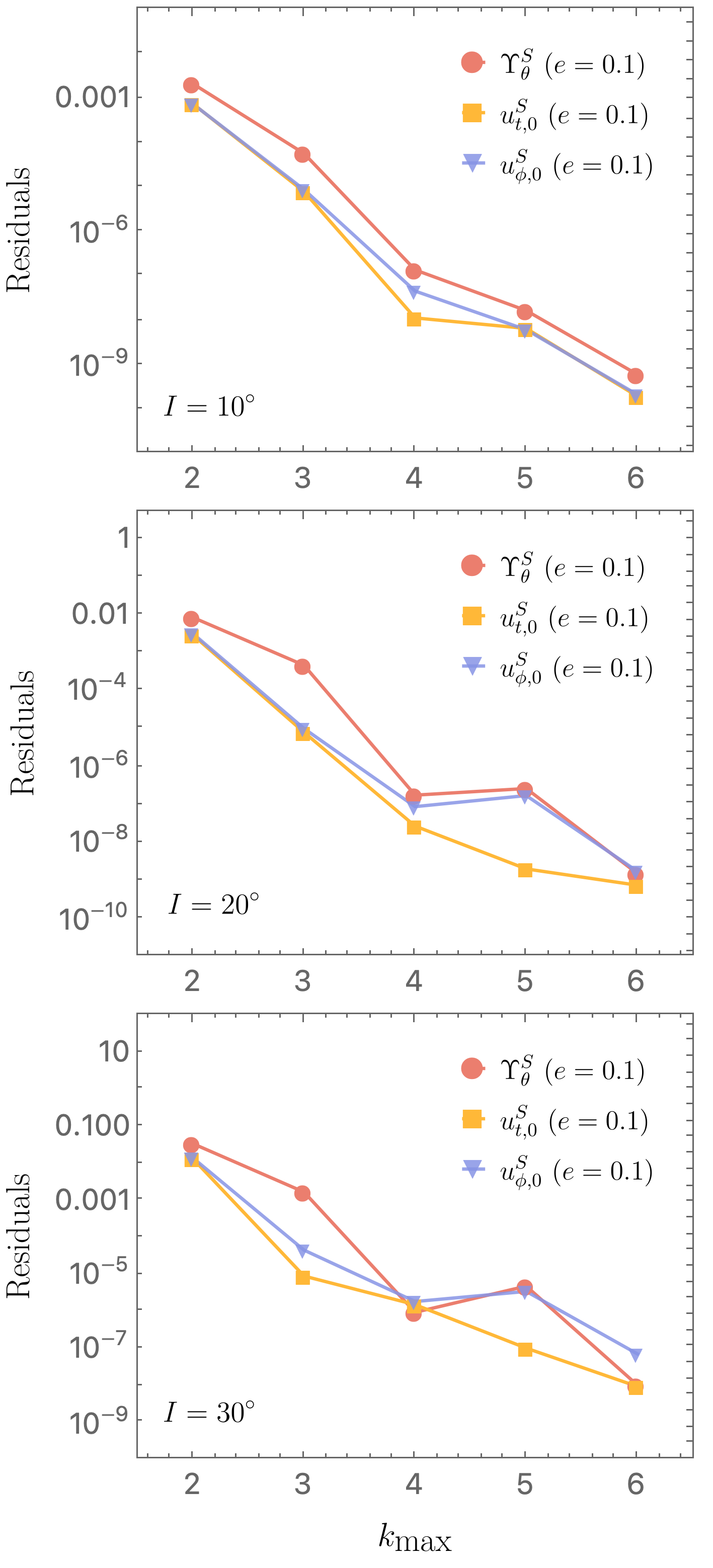}}
\caption{Plot of residuals versus $k_{\text{max}}$ for a generic ($e=0.1$) orbit of an aligned ($s_\parallel=s$) spinning body. We plot $\Upsilon_{\theta}^S$, $u^S_{t,0}$, $u^S_{\phi,0}$ using red circular, orange square and blue triangular markers respectively. As in Fig.\ \ref{fig:residualplotInearlycirc}, we compute the residuals by evaluating the change between subsequent values of $k_\text{max}$ for each of the quantities plotted. Top panel shows $I=10^{\circ}$; middle shows $I=20^{\circ}$; and bottom shows $I=30^{\circ}$. In all cases, $n_{\text{max}}=k_{\text{max}}$, the large black hole has spin parameter $a=0.9M$, and the orbit has $p=10$. \label{fig:residualplotIgeneric}}
\end{figure}

\begin{figure*}
\centerline{\includegraphics[scale=0.73]{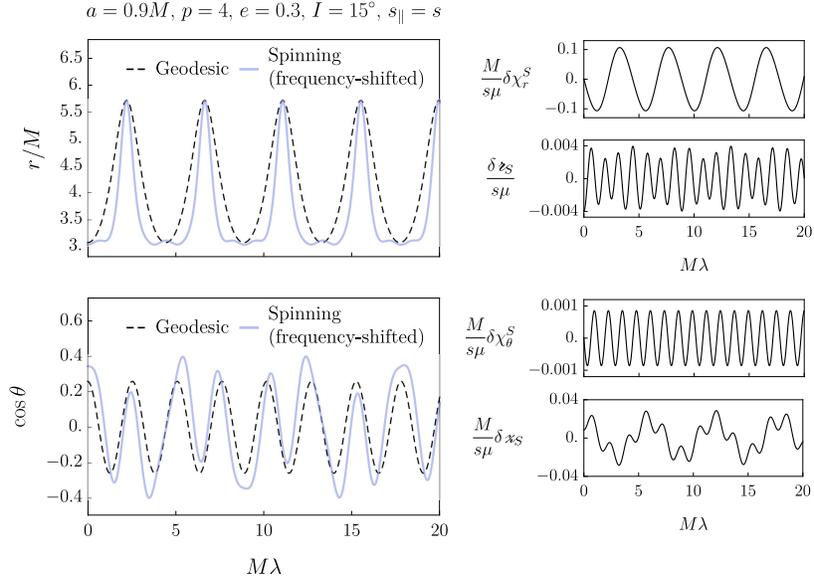}}
\caption{Example of generic orbit motion for an aligned spinning test body around a Kerr black hole with $a=0.9M$. Top left panel shows $r$ versus $\lambda$ for a geodesic (black dashed) and a spinning test body (blue solid) orbit. The radial reference geodesic has $p=4$, $e=0.3$. Note that, in the two left panels, we have used an unphysically high spin $\mu s /M=10$ in order make the spin-curvature effects clearly visible. Also note that the spinning-body orbit has been shifted slightly: its radial frequency $\Upsilon_r= \hat\Upsilon_r + \Upsilon_r^S$ has been replaced with $\hat\Upsilon_r$ and its polar frequency $\Upsilon_\theta= \hat\Upsilon_\theta + \Upsilon_\theta^S$ has been replaced with $\hat\Upsilon_\theta$. This is done so that in the plot the geodesic and the spinning-body orbit remain phase-locked, which helps to illustrate differences in their motion between each turning point. Bottom left panel shows $\cos\theta$ versus $\lambda$ for a geodesic (black dashed) and a spinning test body (blue solid) orbit. The polar reference geodesic has $I=15^\circ$. Again, note that for making this plot, the spinning-body orbit has been shifted slightly: its polar frequency $\Upsilon_\theta= \hat\Upsilon_\theta + \Upsilon_\theta^S$ has been replaced with $\hat\Upsilon_\theta $ for the same reason described above. The right column shows $\delta\chi_r^S$, $\delta \mathcalligra{r}_S$, $\delta\chi_{\theta}^S$ and $\delta \mathcalligra{z}_S$, all versus $\lambda$ and all drawn using black solid lines.  In making these plots, we have used $s_\parallel=s$ and $n_\text{max}=k_\text{max}=3$.  \label{fig:exampleorbits}}
\end{figure*}

\begin{figure*}
\centerline{\includegraphics[scale=0.68]{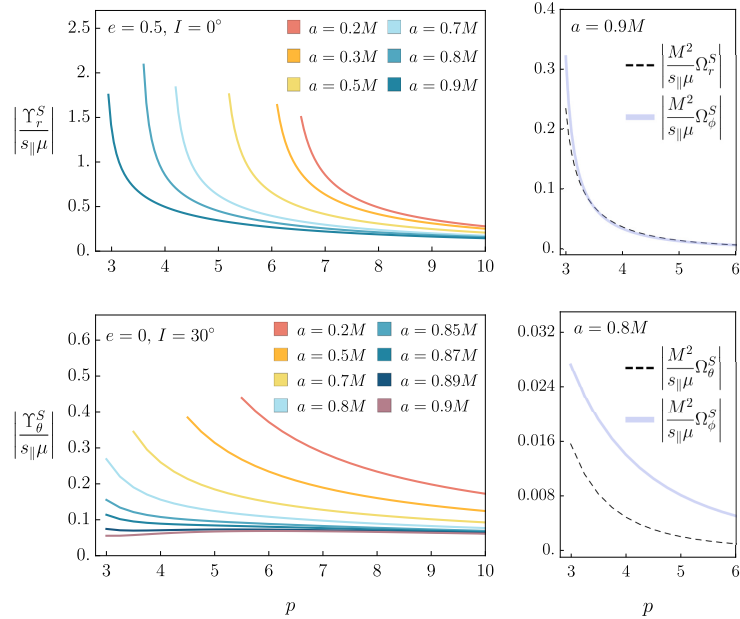}}
\caption{Example of the spin contributions $\Upsilon_r^S$ and $\Upsilon_\theta^S$ to the radial and polar Mino-time frequencies $\Upsilon_r$ and $\Upsilon_\theta$, as well as spin contributions $\Omega_r^S$ and $\Omega_\theta^S$ to the radial and polar coordinate-time frequencies $\Omega_r$ and $\Omega_\theta$. Top left panel shows $\Upsilon_r^S$ versus $p$ with $e=0.5$ and $I=0^{\circ}$ for different values of $a$. Bottom left panel shows $\Upsilon_{\theta}^S$ versus $p$ with $e=0$ and $I=30^{\circ}$ for different values of $a$. Top right panel shows $\Omega_r^S$ (black dashed) and $\Omega_{\phi}^S$ (blue solid) versus $p$ with $a=0.9M$, $e=0.5$ and $I=0^{\circ}$. Bottom right panel shows $\Omega_{\theta}^S$ (black dashed) and $\Omega_{\phi}^S$ (blue solid) versus $p$ with $a=0.8M$, $e=0$ and $I=30^{\circ}$. In making these plots, we have used $n_{\text{max}}=k_{\text{max}}=5$. \label{fig:Upsilonrthetaomega}}
\end{figure*}

\begin{figure}
\centerline{\includegraphics[scale=0.71]{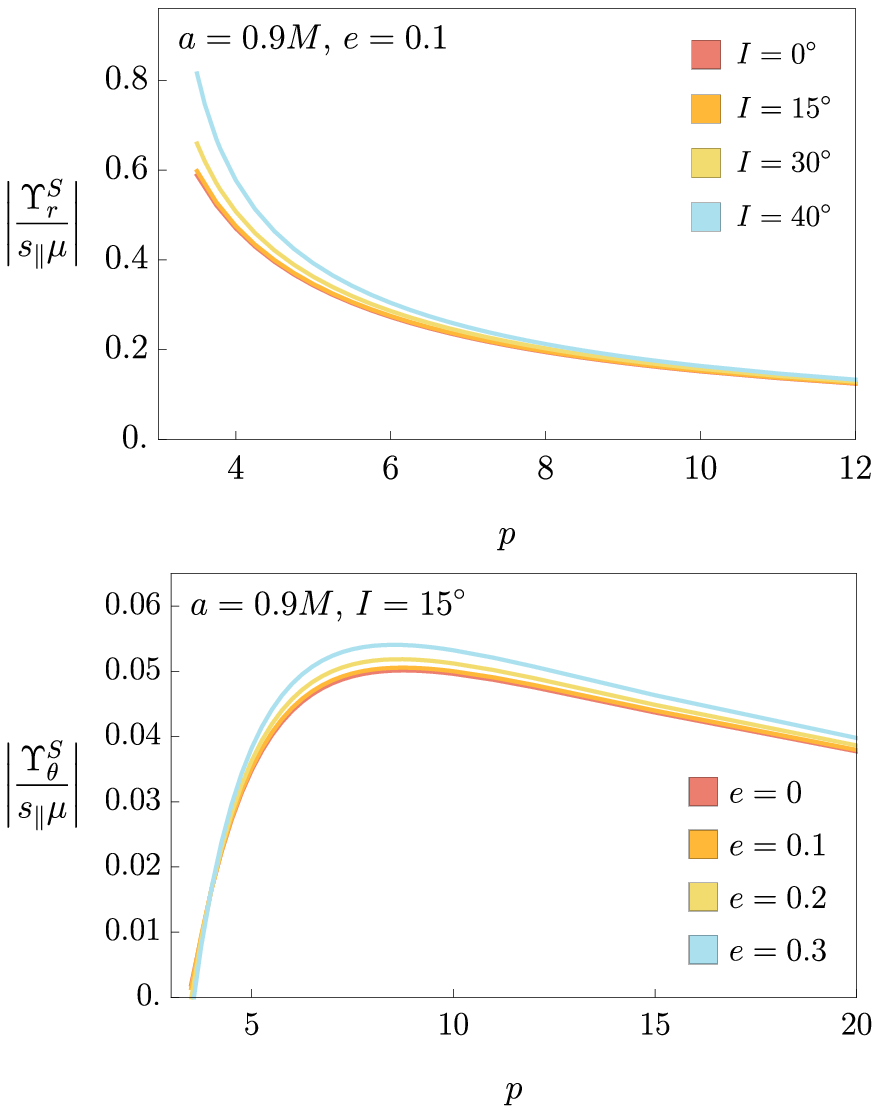}}
\caption{Example of the spin contributions $\Upsilon_r^S$ and $\Upsilon_\theta^S$ to the radial and polar Mino-time frequencies $\Upsilon_r$ and $\Upsilon_\theta$. Top panel shows  $\Upsilon_r^S$ versus $p$ with $e=0.1$ for $I=0^{\circ}$ (red), $I=15^{\circ}$ (orange), $I=30^{\circ}$ (yellow) and $I=40^{\circ}$ (blue). Bottom panel shows $\Upsilon_{\theta}^S$ versus $p$  with $I=15^{\circ}$ for $e=0$ (red), $e=0.1$ (orange), $e=0.2$ (yellow) and $e=0.3$ (blue). In making these plots, we have used $a=0.9M$ and $s_{\parallel}=s$.\label{fig:Upsilongenericplot}}
\end{figure}

In general, the radial and polar motions are coupled and consequently orbits of spinning bodies have radial and polar turning points that vary over the course of the orbit.  This means that positions of the radial turning points depend on $\theta$ and likewise the polar turning points depend on the radial position of the body, as explicitly shown in Ref.\ \cite{Witzany2019_2}. In addition, the turning points depend on the precession phase $\psi_p$ defined in equation (\ref{eq:precphasesol}). Therefore, as in Eqs.\ (\ref{eq:rgenparam1}) -- (\ref{eq:thetagenparam1}), we include the terms $\delta\mathcalligra{r}_S$ and $\delta\mathcalligra{z}_S$ in our parameterization  to capture the modification to the libration range, yielding
\begin{align}
r & =\frac{pM}{1+e\cos\left(w_{r}+\delta\hat{\chi}_r(w_r)+\delta\chi_r^S\right)}+\delta\mathcalligra{r}_S\;,\label{eq:rparamgen}\\
\cos\theta & =\sin I\cos \left(w_{\theta}+\delta\hat{\chi}_{\theta}(w_{\theta})+\delta\chi_{\theta}^S\right)+\delta\mathcalligra{z}_S\;.\label{eq:thetaparamgen}
\end{align}
As described in Sec.\ \ref{sec:RefGeodFreqDom}, the true anomaly angles $\delta\chi_r^S$ and $\delta\chi_{\theta}^S$ contained inside the arguments of the cosines in Eqs.\ (\ref{eq:rparamgen}) and (\ref{eq:thetaparamgen}) consist of purely radial and purely polar oscillations respectively:
\begin{align}
\delta\chi_{r}^{S} & =\sum_{n=-\infty}^{\infty}\delta\chi_{r,n}^{S}e^{-in w_r}\;,\label{eq:deltachirgen}\\
\delta\chi_{\theta}^{S} & =\sum_{k=-\infty}^{\infty}\delta\chi_{\theta,k}^{S}e^{-ikw_{\theta}}\;.\label{eq:deltachithetagen}
\end{align}
Motion that is not purely radial or purely polar is subsumed into the functions $\delta\mathcalligra{r}_S$ and $\delta\mathcalligra{z}_S$.  These quantities are written as Fourier expansions of the form (\ref{eq:Fexpandgeneric}).  For the radial libration variation,
\begin{align}
\delta\mathcalligra{r}_S =\sum_{j=-1}^{1}\sum_{n,k=-\infty}^{\infty}\delta\mathcalligra{r}_{S,jnk}e^{-i(n w_r+k w_{\theta}+j w_s)}\;,\label{eq:deltachir2gen}
\end{align}
where $k$ and $j$ cannot both be zero; for the polar libration variation,
\begin{align}
\delta\mathcalligra{z}_S =\sum_{j=-1}^{1}\sum_{n,k=-\infty}^{\infty}\delta\mathcalligra{z}_{S,,jnk}e^{-i(n w_r+kw_{\theta}+j w_s)}\;,\label{eq:deltachitheta2gen}
\end{align}
where $n$ and $j$ cannot both be zero.

\begin{widetext}
We insert equations (\ref{eq:rparamgen}), (\ref{eq:thetaparamgen}) and (\ref{eq:utuphi}) into (\ref{eq:forcer}) -- (\ref{eq:forcetheta}) and linearize in spin. The radial equation (\ref{eq:forcer}) now has the form 
\begin{align}
\mathcal{F}_r \frac{d^2 \delta\chi_{r}^S}{d \lambda^2}+\mathcal{F}_{\mathcalligra{r}} \frac{d^2 \delta\mathcalligra{r}_S}{d \lambda^2}+\mathcal{G}_r \frac{d \delta\chi_{r}^S}{d \lambda}+\mathcal{G}_{\mathcalligra{r}} \frac{d \delta\mathcalligra{r}_S}{d \lambda}+\mathcal{G}_\theta  \frac{d \delta\chi_{\theta}^S}{d \lambda}+\mathcal{G}_{\mathcalligra{z}}\frac{d \delta\mathcalligra{z}_S}{d \lambda}+H_{r} \delta\chi_{r}^S+\mathcal{H}_{\mathcalligra{r}} \delta\mathcalligra{r}_S\nonumber \\ 
+\mathcal{H}_\theta\delta\chi_{\theta}^S+\mathcal{H}_{\mathcalligra{z}}\delta\mathcalligra{z}_S+\mathcal{I}_{1r} \Upsilon_{r}^S +\mathcal{I}_{1\theta} \Upsilon_{\theta}^S +\mathcal{I}_2 u^S_{t,0} +\mathcal{I}_3 u^S_{\phi, 0}+\mathcal{J} =0\;.\label{eq:radiallinMP5}
\end{align}
As we have seen in earlier expressions, the quantities $\mathcal{F}_r$, $\mathcal{F}_{\mathcalligra{r}}$, $\mathcal{G}_r$,  $\mathcal{G}_{\mathcalligra{r}}$, $\mathcal{G}_\theta$, $\mathcal{G}_{\mathcalligra{z}}$, $\mathcal{H}_r$,  $\mathcal{H}_{\mathcalligra{r}}$,  $\mathcal{H}_\theta$,  $\mathcal{H}_{\mathcalligra{z}}$, $\mathcal{I}_{1r}$, $\mathcal{I}_{1\theta}$, $\mathcal{I}_{2}$, $\mathcal{I}_{3}$ and $\mathcal{J}$ are all functions of known quantities evaluated on geodesics. Eq.\ (\ref{eq:forcetheta}) becomes
\begin{align}
\mathcal{Q}_\theta\frac{d^2 \delta\chi_{\theta}^S}{d \lambda^2}+\mathcal{Q}_{\mathcalligra{z}}\frac{d^2 \delta\mathcalligra{z}_S}{d \lambda^2}+S_r \frac{d \delta\chi_r^S}{d \lambda}+\mathcal{S}_{\mathcalligra{r}} \frac{d \delta\mathcalligra{r}_S}{d \lambda}+\mathcal{S}_\theta\frac{d \delta\chi_{\theta}^S}{d \lambda}+\mathcal{S}_{\mathcalligra{z}}\frac{d \delta\mathcalligra{z}_S}{d \lambda}+\mathcal{T}_r\delta\chi_r^S+\mathcal{T}_{\mathcalligra{r}}\delta\mathcalligra{r}_S\nonumber \\
+\mathcal{T}_\theta\delta\chi_\theta^S+\mathcal{T}_\mathcalligra{z}\delta\mathcalligra{z}_S+\mathcal{U}_{1r}\Upsilon_r^S +\mathcal{U}_{1\theta}\Upsilon_{\theta}^S +\mathcal{U}_{2} u^S_{t,0} +\mathcal{U}_{3}u^S_{\phi, 0}+\mathcal{V}=0\;,\label{eq:polarlinMP5}
\end{align}
where $\mathcal{Q}_\theta$,  $\mathcal{Q}_{\mathcalligra{z}}$, $S_r$, $\mathcal{S}_{\mathcalligra{r}}$, $\mathcal{S}_\theta$, $\mathcal{S}_{\mathcalligra{z}}$, $\mathcal{T}_r$, $\mathcal{T}_{\mathcalligra{r}}$,  $\mathcal{T}_{\theta}$, $\mathcal{T}_\mathcalligra{z}$, $\mathcal{U}_{1r}$, $\mathcal{U}_{1\theta}$, $\mathcal{U}_{2}$, $\mathcal{U}_{3}$ and $\mathcal{V}$ are all functions of known quantities evaluated on geodesics.  We also use $u^{\alpha}u_{\alpha}=-1$ to obtain
\begin{align}
\mathcal{K}_r \frac{d \delta\chi_r^S}{d \lambda}+\mathcal{K}_{\mathcalligra{r}} \frac{d \delta\mathcalligra{r}_S}{d \lambda}+\mathcal{K}_\theta \frac{d \delta\chi_{\theta}^S}{d \lambda}+\mathcal{K}_{\mathcalligra{z}} \frac{d \delta\mathcalligra{z}_S}{d \lambda}+\mathcal{M}_r \delta\chi_r^S+\mathcal{M}_{\mathcalligra{r}} \delta\mathcalligra{r}_S+\mathcal{M}_\theta \delta\chi_{\theta }^S+\mathcal{M}_{\mathcalligra{z}} \delta\mathcalligra{z}_S\nonumber \\
+\mathcal{N}_{1r}\Upsilon_r^S +\mathcal{N}_{1\theta}\Upsilon_{\theta}^S+\mathcal{N}_{2}u^S_{t,0} +\mathcal{N}_{3} u^S_{\phi, 0}+\mathcal{P}=0\;,\label{eq:udotu5}
\end{align}
where $\mathcal{K}_r$, $\mathcal{K}_{\mathcalligra{r}}$, $\mathcal{K}_\theta$, $\mathcal{K}_{\mathcalligra{z}}$, $\mathcal{M}_r$, $\mathcal{M}_{\mathcalligra{r}}$, $\mathcal{M}_\theta$, $\mathcal{M}_{\mathcalligra{z}}$, $\mathcal{N}_{1r}$, $\mathcal{N}_{1\theta}$, $\mathcal{N}_{2}$, $\mathcal{N}_{3}$ and $\mathcal{P}$ are again all known functions evaluated on geodesics. 
\end{widetext}
We describe $\mathcal{F}_r$, $\mathcal{F}_{\mathcalligra{r}}$, $\mathcal{G}_r$, $\mathcal{G}_{\mathcalligra{r}}$, $\mathcal{G}_\theta$, $\mathcal{G}_{\mathcalligra{z}}$, $\mathcal{H}_r$, $\mathcal{H}_{\mathcalligra{r}}$, $\mathcal{H}_\theta$, $\mathcal{H}_{\mathcalligra{z}}$, $\mathcal{I}_{1r}$, $\mathcal{I}_{1\theta}$, $\mathcal{I}_{2}$, $\mathcal{I}_{3}$, $\mathcal{J}$, $\mathcal{Q}_\theta$,  $\mathcal{Q}_{\mathcalligra{z}}$, $S_r$, $\mathcal{S}_{\mathcalligra{r}}$, $\mathcal{S}_\theta$, $\mathcal{S}_{\mathcalligra{z}}$, $\mathcal{T}_r$, $\mathcal{T}_{\mathcalligra{r}}$,  $\mathcal{T}_{\theta}$, $\mathcal{T}_\mathcalligra{z}$, $\mathcal{U}_{1r}$, $\mathcal{U}_{1\theta}$, $\mathcal{U}_{2}$, $\mathcal{U}_{3}$, $\mathcal{V}$, $\mathcal{K}_r$, $\mathcal{K}_{\mathcalligra{r}}$, $\mathcal{K}_\theta$, $\mathcal{K}_{\mathcalligra{z}}$, $\mathcal{M}_r$, $\mathcal{M}_{\mathcalligra{r}}$, $\mathcal{M}_\theta$, $\mathcal{M}_{\mathcalligra{z}}$, $\mathcal{N}_{1r}$, $\mathcal{N}_{1\theta}$, $\mathcal{N}_{2}$, $\mathcal{N}_{3}$, and $\mathcal{P}$ using Fourier expansions of the form (\ref{eq:Fexpandgeneric}).  We provide full expressions for these functions in the \textit{Mathematica} notebook in the Supplemental Material for this paper \cite{SupplementalMaterial}.  We insert these expansions along with (\ref{eq:deltachirgen}), (\ref{eq:deltachithetagen}), (\ref{eq:deltachir2gen}) and (\ref{eq:deltachitheta2gen}) into  (\ref{eq:radiallinMP5}), (\ref{eq:polarlinMP5}) and (\ref{eq:udotu5}). We then solve for the unknown variables $\delta\chi_r^S$, $\delta\mathcalligra{r}_S$, $\delta\chi_{\theta}^S$, $\delta\mathcalligra{z}_S$, $\Upsilon_r^S$, $\Upsilon_{\theta}^S$, $u_{t}^S$ and $u_{\phi}^S$. This frequency-domain approach therefore naturally allows us to compute the first-order-in-spin corrections to the orbital frequencies $\Upsilon_r^S$ and $\Upsilon_{\theta}^S$ for totally generic orbits of spinning particles. 

In Fig.\ \ref{fig:residualplotIgeneric}, we see how $\Upsilon_{\theta}$  converges to its true values as $n_{\text{max}}$ and $k_{\text{max}}$ increase for a reference geodesic that is both inclined and eccentric. In Sec.\ \ref{sec:circinclalign}, the convergence of $\Upsilon_{\theta}$ for nearly circular ($e=0$) orbits is plotted in Fig.\ \ref{fig:residualplotInearlycirc}. At all inclinations, at $n_{\text{max}}=2$, the residuals are smaller for the $e=0$ reference orbit than for the slightly eccentric $e=0.1$ reference orbit. For the smallest inclination $I=10^{\circ}$ (top panel), this difference holds for all $n_{\text{max}}$; the quantities corresponding to a reference geodesic that is both eccentric and inclined (Fig.\ \ref{fig:residualplotIgeneric}) all converge slower than those corresponding to a reference geodesic that is nearly circular and inclined (Fig.\ \ref{fig:residualplotInearlycirc}). However, as the inclination is increased, the difference in the rate of convergence between the eccentric and circular cases decreases. At the highest inclination, $I=30^{\circ}$, they converge at roughly the same rate. 

Fig.\ \ref{fig:exampleorbits} shows an example of $r$ and $\theta$ for a generic spinning-body orbit, in addition to the functions $\delta\chi_r^S$, $\delta\chi_{\theta}^S$, $\delta\mathcalligra{r}_S$ and $\delta\mathcalligra{z}_S$ which go into constructing the orbit's $r$ and $\theta$.  In the two left-hand panels of Fig.\ \ref{fig:exampleorbits}, the reference geodesic orbit associated with this spinning-body orbit is overplotted with a dotted black curve; both the radial and polar frequencies associated with the spinning-body orbit are shifted so that it remains phase-locked with the geodesic reference orbit.  In addition, $\mu s/M$ has been chosen to have an unphysically large value of 10 in order to clearly show the effect of spin-curvature coupling on the shape of the orbit. 

Figure \ref{fig:Upsilonrthetaomega} shows how $\Upsilon_r^S$ varies with $p$ for nearly equatorial eccentric orbits, and likewise how $\Upsilon_\theta^S$ varies with $p$ for nearly circular inclined orbits.  Notice that the spin corrections to the polar Mino-time frequencies $\Upsilon^S_\theta$ (bottom left panel) have a different dependence on $p$ compared to the radial Mino-time frequencies $\Upsilon^S_r$ (top left panel).  For all spins, we see that the radial correction $\Upsilon^S_r$ increases rapidly near the last stable orbit (LSO).  For small values of $a$, the behavior of $\Upsilon^S_\theta$ is similar, increasing as orbits approach the LSO, albeit with a shallower slope.  However, for large $a$ ($a>0.8M$), a different trend emerges.  For $a=0.85M$ and $a=0.87M$ the curve flattens, with a slight uptick as it approaches the LSO; for $a=0.89M$ and $a=0.9M$, the curve reaches a maximum and begins to trend downwards very close to the LSO, as can be seen in the bottom left panel of Fig.\ \ref{fig:Upsilonrthetaomega}.  The dependence of the frequency corrections on $a$ is fairly similar for both $\Upsilon^S_r$ and $\Upsilon^S_\theta$: In both cases, at fixed $p$, the frequency correction is larger for the smaller value of $a$. Fig.\ \ref{fig:Upsilonrthetaomega} displays coordinate-time frequency corrections $\Omega_r$, $\Omega_{\theta}$ and $\Omega_{\phi}$ for an equatorial orbit (top right panel) and an inclined orbit (bottom right panel).

Fig.\ \ref{fig:Upsilongenericplot} shows how the corrections to the radial $\Upsilon^S_r$ and polar $\Upsilon^S_{\theta}$ Mino-time frequencies vary with $p$, $e$ and $I$ when the reference geodesic is both inclined and eccentric. In Fig.\ \ref{fig:Upsilongenericplot}, we see similar trends to those in Fig.\ \ref{fig:Upsilonrthetaomega}. In the bottom panel of Fig.\ \ref{fig:Upsilongenericplot}, $\Upsilon^S_{\theta}$ increases with decreasing $p$ until it reaches a maximum and then begins to decrease as $p$ approaches the LSO.  Increasing the eccentricity of the orbit shifts the maximum $\Upsilon^S_{\theta}$  to a higher value. In the top panel of Fig.\ \ref{fig:Upsilongenericplot}, $\Upsilon^S_r$ increases with decreasing $p$. Increasing the inclination angle of the orbit leads to a more rapid increase in $\Upsilon^S_r$ as $p$ approaches the LSO. 

\section{Summary and future work}
\label{sec:summary}

In this paper, we present a frequency-domain approach for precisely computing the orbits of spinning bodies.  This extends the work presented in our companion paper \cite{Paper1} by considering completely generic orbits with arbitrarily oriented spin, going beyond the equatorial and nearly equatorial orbits discussed previously. In Sec.\ \ref{sec:spinningbodyorbits}, we outline our perturbative approach to studying spinning-body dynamics both qualitatively and quantitatively, and in Sec.\ \ref{sec:freqdomdescrip}, we describe how we compute spinning-body orbits in the frequency-domain. In Sec.\ \ref{sec:results}, we discuss the results we obtain using frequency-domain methods; in particular, we compute the corrections to the radial $\Upsilon_r^S$ and polar $\Upsilon_\theta^S$ frequencies due to the spin of the orbiting body.

There are several future avenues we plan to explore related to this work. First, we aim to study the role played by nonlinear-in-spin terms near resonance in pushing the spinning-body dynamics from integrable to chaotic via the KAM theorem; this would extend the preliminary investigation in Ref.\ \cite{Ruangsri2016}. Second, we are working on incorporating secondary spin into gravitational waveform models using an osculating geodesic scheme \cite{Pound2008,Gair2011}. For example, this method has already been applied to produce spinning-body inspirals for a Schwarzschild background in Ref.\ \cite{Warburton2017}.  Our goal is to build a framework for completely generic adiabatic inspirals of spinning bodies.

In addition, we aim to systematically explore and present the orbital frequencies obtained in this work. First, we want to explicitly demonstrate that the frequencies obtained in Ref.\ \cite{Witzany2019_2} are entirely equivalent to those presented in this work. We explicitly show the equivalence of the two approaches for the equatorial spin-aligned case in App.\ \ref{sec:comparWitzany} and we intend to extend this comparison to include frequencies associated with completely generic orbits. Second, we plan to compare with Post-Newtonian results based on the analysis in Refs.\ \cite{Tanay2021_1} and \cite{Tanay2021_2} as another validity check of our results. A catalog of these frequencies and how they vary with the parameters describing orbits and the small body's spin orientation is likely to be of use as waveform models for large mass ratio systems are developed and incorporated in gravitational-wave measurement pipelines.

\section*{Acknowledgements}

This work has been supported by NASA ATP Grant 80NSSC18K1091, and NSF Grant PHY-1707549 and PHY-2110384. We are very grateful to Vojt\v{e}ch Witzany for reading through a draft of this paper and providing helpful feedback.

\appendix

\section{Reference geodesics}
\label{sec:referencegeodesics}

There are different mappings that can be constructed from the triplet of constants $(p,e,I)$ defining a geodesic (i.e., the ``reference" geodesic) to a particular spinning-body orbit. The choice of reference geodesic we use in this article is discussed in Sec.\ \ref{sec:RefGeodFreqDom}. In brief, we find spinning-body orbits for which the ``purely radial" and ``purely polar" components of the motion have the same turning points as the reference geodesic; see Sec.\ \ref{sec:RefGeodFreqDom} for mathematical details.  However, there are other physically equivalent mappings which can be used instead and may be particularly useful in certain circumstances. We outline three approaches that have appeared in the literature below.

\subsection{Reference geodesic has the same turning points as the spinning-body orbit}
\label{sec:refgeod1}
The definition of reference geodesic we use in this work is most similar to that used by Mukherjee et al.\ in Ref.\ \cite{Mukherjee2019} and Skoupy et al. in Ref.\ \cite{Skoupy2021}. In Refs.\ \cite{Mukherjee2019} and \cite{Skoupy2021}, they study eccentric equatorial orbits where the spin is aligned, and in this case the reference geodesic has the same radial \textit{turning points} as the spinning-body orbit under consideration. In our approach, we generalize this for generic orbital configurations and misaligned small-body spin: The purely radial and purely polar parts of the spinning-body motion are constrained to have the same libration range as the corresponding reference geodesic. Complementary to this, there are additional corrections to the libration range due to motion that is not purely radial, or purely polar (see Sec.\ \ref{sec:RefGeodFreqDom}). For example, if the reference geodesic is equatorial, the corresponding spinning-body orbit is not equatorial except in the aligned spin case. Instead, it lies $\mathcal{O}(S)$ out of the equatorial plane. An example of a reference geodesic with the same radial turning points as the corresponding spinning-body orbit is shown in Fig.\ \ref{fig:refgeod}(a).

\subsection{Reference geodesic has the same initial conditions as the spinning-body orbit}
\label{sec:refgeod2}
In the analyses by Bini et al., the reference geodesic is defined as the geodesic that has the same \textit{initial conditions} as the corresponding spinning-body orbit \cite{Bini2011_1,Bini2011_2}. Work by Mashoon et al.\ takes a similar approach \cite{Mashhoon2006}. For example, in  Ref.\ \cite{Bini2011_1}, analytic expressions for a spinning-body orbit with the same initial position and 4-velocity as a circular equatorial reference geodesic is obtained; this calculation can represent a scenario where spin-curvature force is ``turned on" at a certain point along a geodesic orbit and subsequent spinning-body motion is computed. An example of a reference geodesic with the same initial conditions as the corresponding spinning-body orbit is presented in Fig.\ \ref{fig:refgeod}(b).

\subsection{Reference geodesic has the same constants of motion as the spinning-body orbit}
\label{sec:refgeod3}
\begin{figure}
\centerline{\includegraphics[scale=0.7]{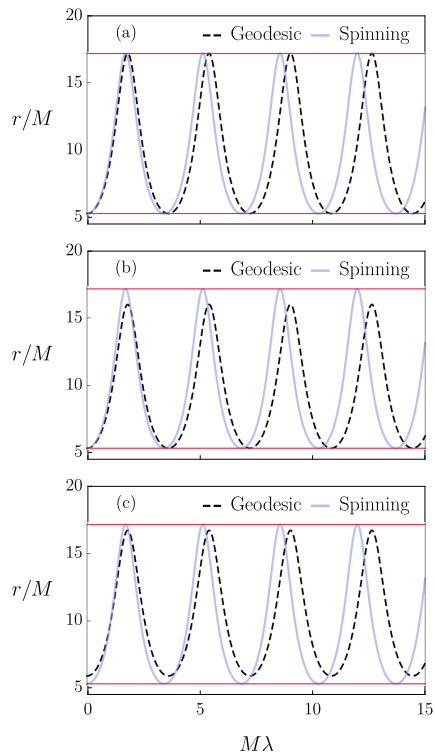}}
\caption{Example of radial motion for an aligned, spinning body in an equatorial orbit around a non-rotating black hole ($a=0$). All panels show $r$ versus $\lambda$ for a spinning body (blue) and corresponding reference geodesic (black, dashed) orbit. Radial turning points, corresponding to $p=8.13721$, $e=0.525726$, of the spinning body's orbit are shown by the solid red lines. Different choices of reference geodesic for the same spinning-body orbit are shown in (a), (b) and (c). \textit{Top.} (a) The spinning-body orbit and reference geodesic have the same turning points.  \textit{Middle.} (b) The spinning-body orbit and reference geodesic have the same initial conditions. \textit{Bottom.} (c) The spinning-body orbit and reference geodesic have the same constants of motion. In making these plots, we have used $s_{\parallel}=s$ and $\mu s/M=0.05$. \label{fig:refgeod}}
\end{figure}

 In the analysis by Witzany in Ref.\ \cite{Witzany2019_2}, the ``fiducial" geodesic is taken to be the geodesic with the same \textit{constants of motion} as the spinning-body orbit, modulo a $-2as_{\parallel}\text{sgn}(L_z-aE)$ correction to the definition of the Carter constant $K$. The inclusion of the $-2as_{\parallel}\text{sgn}(L_z-aE)$ term in the choice of fiducial mapping ensures that the formulae for the turning point corrections presented in Eq.\ (48) of Ref.\ \cite{Witzany2019_2} are finite for motion in the equatorial plane. The turning point spin-corrections corresponding to those constants of motion are then computed and used to parameterize the orbital motion. An example of a reference geodesic with the same constants of motion as the corresponding spinning-body orbit is presented in Fig.\ \ref{fig:refgeod}(c). See App.\ \ref{sec:comparWitzany} for a detailed discussion of the approach in Ref.\ \cite{Witzany2019_2} and an explicit comparison with our formulation for the case of equatorial, aligned-spin orbits in a Schwarzschild background. 
 
\section{Comparison with Witzany, 2019}
\label{sec:comparWitzany}
In Ref.\ \cite{Witzany2019_2}, Witzany outlines an approach for obtaining the equations of motion for spinning bodies to first-order in spin using the Hamilton-Jacobi equation. This approach yields the equations of motion Eqs. 46(a) -- (c) in Ref.\ \cite{Witzany2019_2} which we reproduce here:
\begin{align}
    \frac{dr}{d\lambda}&=\pm \Delta \sqrt{w_r^{\prime 2}-e_{0r} e^{\kappa}_{C;r} e_{\kappa B}\tilde{s}^{CD}},\label{eq:drdlambdaWitzany} \\ 
     \frac{d\theta}{d\lambda}&=\pm \sqrt{w_{\theta}^{\prime 2}-e_{0\theta}e^{\kappa}_{C;\theta}e_{\kappa B}\tilde{s}^{CD}},\label{eq:dthetadlambdaWitzany}\\
     \frac{d\psi_p}{d\lambda}&=\sqrt{\hat{K}}\left(\frac{(r^2+a^2)\hat{E}-a\hat{L}_z}{\hat{K}+r^2}+a\frac{\hat{L}_z-a(1-z^2)\hat{E}}{\hat{K}-a^2z^2}\right),\label{eq:paralleltransport2}
\end{align}
where the tetrad $e^{\kappa}_C$ is the parallel transported tetrad given by Eqs.\ (\ref{eq:tetradleg1}) -- (\ref{eq:tetradleg2}) in Sec.\ \ref{sec:paralleltransport}. Here $B$, $C$ and $D$ are labels for the tetrad legs. Note that $\psi_p$ in Eq.\ (\ref{eq:paralleltransport2}) is denoted $\phi$ in Ref.\ \cite{Witzany2019_2}; Eq.\ (\ref{eq:paralleltransport2}) is identical to Eq.\ (\ref{eq:precphaseeqn}). The expressions for $\tilde{s}_{CD}$ are underneath Eq.\ (33) in Ref.\ \cite{Witzany2019_2}. From Eqs.\ (\ref{eq:drdlambdaWitzany}) -- (\ref{eq:paralleltransport2}), we can find the turning points of the equations of motion using the condition that the 4-velocities vanish:
\begin{align}
(w^{\prime}_y)^2-e_{0y}e^\kappa_{C;y}e_{D\kappa}s^{CD}=0\label{eq:turningpointcondition},
\end{align}
where $y=r$, $\theta$. Using condition (\ref{eq:turningpointcondition}), Witzany derives analytical expressions for the corrections to the turning points due to the small body's spin. These expressions can be found in Eqs.\ 48(a) -- (f) of Ref.\ \cite{Witzany2019_2} and apply for fully generic orbits in the first order in $S$ limit. 

\subsection{Description of the two approaches}

The framework used in Ref.\ \cite{Witzany2019_2} is an alternative method for calculating spinning-body orbital frequencies $\Upsilon_r$ and $\Upsilon_\theta$. In an approach analogous to that used by Carter in Ref.\ \cite{Carter1968}, Witzany uses the Hamilton-Jacobi equation to obtain expressions for $dr/d\lambda$ and $d\theta/d\lambda$, yielding Eqs.\ (\ref{eq:drdlambdaWitzany}) -- (\ref{eq:paralleltransport2}). The Mino-time frequencies $\Upsilon_r$ and $\Upsilon_\theta$ are then calculated by integrating these velocities with respect to angle-type coordinates; this procedure is in turn analogous to that used in Refs. \cite{DrascoHughes2004, FujitaHikida2009} to compute geodesic Mino-time frequencies.  The approach we use in this article is to solve the Mathisson-Papapetrou equations (\ref{eq:mp1linear}) -- (\ref{eq:mp2linear}) directly in the frequency-domain. We introduce a frequency correction explicitly into our parameterization and solve for it as one of the unknowns in a linear-algebraic system.

The orbital motion of the spinning body is parameterized differently in the two descriptions. In Ref.\ \cite{Witzany2019_2}, analytic expressions for the corrections to the turning points are obtained using the aforementioned Eqs.\ (\ref{eq:drdlambdaWitzany}) -- (\ref{eq:dthetadlambdaWitzany}). The spinning body's motion is then parameterized in terms of these analytic expressions for the turning points. In our analysis, we do not have explicit expressions for turning point corrections built into our parameterization. Instead, we divide the corrections to the motion of the spinning body into two categories: We include corrections which do not alter the libration range relative to the reference geodesic ($\delta \chi_r^S$, $\delta \chi_\theta^S$), as well as corrections which do modify the libration range ($\delta\mathcalligra{r}_S$, $\delta\mathcalligra{z}_S$). 

In summary, in Ref.\ \cite{Witzany2019_2}, the constants of motion $(E, L_z, K)$ associated with a certain geodesic (called the ``fiducial geodesic", as discussed in App.\ \ref{sec:refgeod3}) are selected, and the turning point corrections for the corresponding spinning-body orbit with the same constants of motion are computed (modulo a $-2as_{\parallel}\text{sgn}(L_z-aE)$ adjustment to $K$), whereupon the frequency corrections can be obtained. Contrastingly, in our framework, we begin by choosing the turning points $({p, e, I})$ for a particular reference geodesic. We then compute the spinning-body orbit which has purely radial and purely polar motion constrained to match the turning points of the reference geodesic. The concomitant frequency corrections and constants of motion for that orbit can then be computed. We show below that our method is consistent with Ref.\ \cite{Witzany2019_2} for orbits in the equatorial plane with aligned small body spin\footnote{As mentioned in Ref.\ \cite{Witzany2019_2}, Witzany conducted a similar consistency check using the effective potential given by Tod et al.\ \cite{1976Tod} and Hackmann et al.\ \cite{Hackmann2014}.}; we leave a detailed comparison of the frequency corrections for fully generic orbits for future work.

\subsection{Numerical comparison of the two approaches}

We compare between the method described in this paper and that presented in Ref.\ \cite{Witzany2019_2} by evaluating the expressions for the radial turning point corrections. For equatorial orbits of a small body with aligned spin and $a=0$, Eqs.\ 48(a) -- (f) in Ref.\ \cite{Witzany2019_2} become:
\begin{align}
    \mathcal{G}&=L_z^S E^S r^2\;,\ \ \ \mathcal{I}=\frac{d}{dr}\left(\frac{(E^Sr^2)^2}{\Delta}-r^2\right)\;,\\
    e_{0y}e_{C\kappa;y}e^\kappa_D&=2\frac{E^S L_z^S \left[r(L_z^S)^2-Mr^2-3M(L_z^S)^2\right]}{r\left[(L_z^S)^2+r^2\right](r-2M)^2}\;,\\
    \delta r&=\left.- s\mu\frac{2\mathcal{G}+\Delta\left[(L_z^S)^2+r^2\right]e_{0y}e_{C\kappa;y}e^\kappa_D}{\mathcal{I}\left[(L_z^S)^2+r^2\right]}\right\rvert_{r_{gt}}\;,
\end{align}
where $\delta r$ is the radial turning point correction evaluated at the fiducial geodesic turning points, which are denoted $r_{gt}$. This reduces to a simple expression for $\delta r$:
\begin{align}
\delta r=\left.s\mu\frac{E^S L_z^S (r-2M)(r-3M)}{r\left[(r-2M)^2-r(E^S)^2(r-3M)\right]}\right\rvert_{r=r_{gt}}\;.\label{eq:deltar}
\end{align}
Note that $E^S$ and $L_z^S$ here are the energy and angular momentum of the spinning-body orbit. As discussed in App.\ \ref{sec:refgeod3}, the fiducial geodesic is the geodesic orbit that has the same energy and angular momentum as the spinning-body orbit we are considering, i.e., $\hat{E}_{\text{fid}}=E^S$ and $\hat{L}_{\text{z,fid}}=L_z^S$.  Eq.\ (\ref{eq:deltar}) is evaluated at the turning points of the fiducial geodesic, $r_{gt1}$ and $r_{gt2}$, and gives the correction to these turning points $\delta r(r_{gt1})$ and $\delta r(r_{gt2})$ due to the spin of the small body. 

\subsubsection{Procedure for computing turning points}
\label{sec:procedure}
As discussed in Appendix \ref{sec:referencegeodesics}, the approach in Ref.\ \cite{Witzany2019_2} is to consider a fiducial geodesic which has the same constants of motion as the spinning-body orbit; this fiducial geodesic has turning points given by $r_{gt1}$ and $r_{gt2}$. The turning point corrections are then computed using Eqs.\ 48(a) -- (f) in Ref.\ \cite{Witzany2019_2}.
\begin{enumerate}
    \item We begin with the constants of motion for a spinning-body orbit with semi-latus rectum $p$ and eccentricity $e$. The energy  $E^S$ and angular momentum $L_z^S$ corresponding to this choice of $p$ and $e$ are given by 
\begin{align}
    E^S=\hat{E}+\delta E^S\;, \ \ \ L_z^S=\hat{L}_z+\delta L_z^S\;,\label{eq:deltaEdeltaL}
\end{align} 
where expressions for $\hat{E}$, $\hat{L}_z$, $\delta E^S$ and $\delta L_z^S$ are given by Eqs.\ (B15), (B16) and (B17) of Ref.\ \cite{Paper1}. We reproduce these equations below:
\begin{align}
    \hat{E}&=\sqrt{\frac{(p-2)^2-4e^2}{p(p-3-e^2)}}\;, \ \ \ \hat{L}_z=\frac{pM}{\sqrt{p-3-e^2}},\label{eq:hatEhatLSchw}\\
    \delta E^S&=-\frac{s\mu}{M}\frac{(1-e^2)^2}{2p(p-3-e^2)^{3/2}}\;, \label{eq:ESexactineSchw}\\
\delta L_z^S&=s\mu\frac{(2p-9-3e^2)\sqrt{(p-2)^2-4e^2}}{2p^{1/2}(p-3-e^2)^{3/2}}\;.\label{eq:LzSexactineSchw}
\end{align}
\item By using $\hat{E}_{\text{fid}}=E^S$ and $\hat{L}_{\text{z,fid}}=L_z^S$ and inverting equations (\ref{eq:hatEhatLSchw}), we can find expressions for the semi-latus rectum $p_{\text{fid}}$ and eccentricity $e_{\text{fid}}$ of a  geodesic orbit, given $E^S$ and $L_z^S$. Notice that these are \textit{not} that same as semi-latus rectum and eccentricity of the spinning-body orbit --- they are the semi-latus rectum and eccentricity corresponding to a geodesic orbit that has the same energy $E^S$ and angular momentum $L_z^S$ as the spinning-body orbit we are considering. 
\item Then, the fiducial turning points can be found, using:
\begin{align}
r_{gt1}=\frac{p_{\text{fid}}M}{1-e_{\text{fid}}}\;, \ \ \ r_{gt2}=\frac{p_{\text{fid}}M}{1+e_{\text{fid}}}\;.\label{eq:fidTP}
\end{align}
\item Next, we evaluate Eq.\ (\ref{eq:deltar}) to find $\delta r$ at each of these fiducial turning points: $\delta r(r_{gt1})$ is the correction to the fiducial apastron and $\delta r(r_{gt2})$ is the correction to the fiducial periastron.
We add these corrections to find the spin-correction turning points:  
\begin{align}
    r_{st1}&=r_{gt1}+\delta r(r_{gt1})\;,\ \ \ r_{st2}=r_{gt2}+\delta r(r_{gt2})\;.
\end{align}
\item We can convert these turning points $r_{st1}$ and $r_{st1}$ into semi-latus rectum $p$ and eccentricty $e$ of the spinning-body orbit using:
\begin{align}
pM=\frac{2r_{st1}r_{st2}}{r_{st1}+r_{st2}}\;, \ \ e=\frac{r_{st1}-r_{st2}}{r_{st1}+r_{st2}}\;.\label{eq:newpe}
\end{align}
\end{enumerate}

\subsubsection{Numerical example}
We follow the procedure outlined in Sec.\ \ref{sec:procedure} with a specific numerical example. For this example case, we already know the turning points of the radial motion and we verify that the turning points computed using Eq.\ (\ref{eq:deltar}) are consistent. Consider a spinning-body orbit with small-body spin $\mu s=0.001M$, semi-latus rectum $pM=7M$ and eccentricity $e=0.4$.
\begin{enumerate}
\item From Eq.\ (\ref{eq:deltaEdeltaL}), this orbit has $E^S=0.951965$ and $L_z^S=3.57273 M$. 
\item Using  $\hat{E}_{\text{fid}}=E^S$ and $\hat{L}_{\text{z,fid}}=L_z^S$ and inverting equations (\ref{eq:hatEhatLSchw}), we find that $p_{\text{fid}}=7.05356$ and $e_{\text{fid}}=0.394709$. 
\item Next, we find the fiducial turning points $r_{gt1}$ and $r_{gt2}$ using Eqs.\ (\ref{eq:fidTP}); they are $r_{gt1}=11.6532M$ and $r_{gt1}=5.05737M$. 
\item Then, we find that $\delta r(r_{gt1})=0.0135323M$ and $\delta r(r_{gt2})=-0.0517264 M$. The spinning-body turning points are $r_{st1}=11.6667M$ and $r_{st2}=5.00564M$.
\item  The spinning-body $p$ and $e$ are found using Eq.\ (\ref{eq:newpe}): $p=7.00553$ and $e=0.399527$. We have recovered the expected $p$ and $e$ for this orbit.
\end{enumerate}

\bibliographystyle{unsrt}
\bibliography{FreqDomSpinningBody}
\end{document}